\documentclass[final,3p,times,authoryear]{elsarticle}

\usepackage{amssymb}
\usepackage{amsmath}
\usepackage{multirow}
\usepackage{subcaption}
\usepackage{float}

\usepackage[utf8]{inputenc}
\usepackage{xcolor}
\usepackage{makecell}
\setcitestyle{square,comma,numbers,sort&compress}

\journal{NIM A}

\begin{document}

\begin{frontmatter}



\title{Calibration of the ComPair Balloon Instrument} 

\author[a,b]{Nicholas Kirschner\corref{cor1}}
\cortext[cor1]{Corresponding author.}
\ead{nkirschner@gwu.edu}
\author[c,d]{Zachary Metzler}
\author[d]{Lucas D. Smith}
\author[b]{Carolyn Kierans}
\author[b]{Regina Caputo}
\author[b]{Nicholas Cannady}
\author[b,c,d]{Makoto Sasaki}
\author[g]{Daniel Shy}
\author[b,c,e]{Priyarshini Ghosh}
\author[f]{Sean Griffin}
\author[g]{J. Eric Grove}
\author[b]{Elizabeth Hays}
\author[b,e]{Iker Liceaga-Indart}
\author[h]{Emily Kong}
\author[b]{Julie McEnery}
\author[b]{John Mitchell}
\author[b,c,d]{A. A. Moiseev}
\author[i]{Lucas Parker}
\author[b]{Jeremy S. Perkins}
\author[g]{Bernard Phlips}
\author[b,j]{Adam J. Schoenwald}
\author[g]{Clio Sleator}
\author[k]{Jacob Smith}
\author[b,l]{Janeth Valverde}
\author[b,c,e]{Sambid Wasti}
\author[g]{Richard Woolf}
\author[g]{Eric Wulf}
\author[b,c,l]{Anna Zajczyk}

\affiliation[a]{organization={The Department of Physics, The George Washington University}, addressline={725 21st St NW}, city={Washington}, state={DC}, postcode={20052}, country={USA}}
\affiliation[b]{organization={NASA Goddard Space Flight Center}, city={Greenbelt}, state={MD},postcode={20771}, country={USA}}
\affiliation[c]{organization={Center for Research and Exploration in Space Science and Technology, NASA/GSFC}, city={Greenbelt}, state={MD}, postcode={20771}, country={USA}}
\affiliation[d]{organization={University of Maryland at College Park}, city={College Park}, state={MD}, postcode={20742}, country={USA}}
\affiliation[e]{organization={Catholic University of America}, addressline={620 Michigan Ave NE}, city={Washington}, state={DC}, postcode={20064}, country={USA}}
\affiliation[f]{organization={Wisconsin IceCube Particle Astrophysics Center, University of Wisconsin-Madison}, addressline={222 W Washington Ave Unit 500}, city={Madison}, state={WI}, postcode={53703}, country={USA}}
\affiliation[g]{organization={Space Science Division, Naval Research Laboratory}, addressline={4555 Overlook Ave SW}, city={Washington}, state={DC},
postcode={20375}, country={USA}}
\affiliation[h]{organization={Technology Service Corporation}, city={Arlington}, state={VA}, postcode={22202}, USA}
\affiliation[i]{organization={Los Alamos National Laboratory}, city={Los Alamos}, state={NM}, postcode={87544}, country={USA}}
\affiliation[j]{organization={University of Maryland, Baltimore County}, addressline={1000 Hilltop Circle}, city={Baltimore}, state={MD}, postcode={21250}, country={USA}}
\affiliation[k]{organization={George Mason University, resident at the Naval Research Laboratory}, addressline={4555 Overlook Ave SW}, city={Washington}, state={DC}, postcode={20375},country={USA}}
\affiliation[l]{organization={Center for Space Sciences and Technology, University of Maryland, Baltimore County}, addressline={1000 Hilltop Circle}, city={Baltimore}, state={MD}, postcode={21250}, country={USA}}

\begin{abstract}
ComPair, the prototype of the All-sky Medium Energy Gamma-ray Observatory (AMEGO) mission concept, is a combined Compton imager and pair production telescope. It consists of four subsystems: a double-sided silicon strip detector (DSSD) Tracker, a virtual Frisch-grid cadmium zinc telluride (CZT) Low Energy Calorimeter, a cesium iodide (CsI) High Energy Calorimeter, and a plastic scintillator Anti-Coincidence Detector (ACD) to reject the charged particle background. These subsystems work together to reconstruct events, by tracking the locations and energies of gamma-ray scatters and pair production events. To quantify ComPair's scientific capabilities prior to a balloon launch in 2023, calibrations were performed to benchmark the instrument's performance in terms of angular resolution, energy resolution, and effective area. In this paper we provide an overview of the ComPair instrument and detail the calibration campaign. Finally, we compare our results to the expected performance based on simulations.

\end{abstract}



\end{frontmatter}



\section{Introduction}
\label{sec1}
ComPair is the prototype instrument for the All-sky Medium Energy Gamma-ray Observatory (AMEGO), which aims to cover the infamous MeV gap \cite{AMEGO}. AMEGO, a probe class mission concept, was designed to advance MeV astrophysics in three areas: time-domain and multimessenger astrophysics enabled by its wide field of view, astrophysical jets and magnetospheres with gamma-ray polarization measurements, and element formation through nuclear line spectroscopy.

Below $\sim$10 MeV, gamma-ray photons predominantly Compton scatter in the detector \cite{Kierans_2022}. Above $\sim$10 MeV, gamma-rays often interact with the detector to produce electron-positron pairs \cite{Thompson-Moiseev_2022}, therefore, to be sensitive in the MeV regime, AMEGO must be able to measure both Compton and pair production events. It achieves this through the use of a 3-subsystem detector stack consisting of a silicon tracker and two calorimeters. This large volume stack enables AMEGO to detect and characterize gamma-rays that undergo both Compton scattering and pair production, by measuring the position and energy deposits from ionizing secondary particles. 

ComPair, the prototype of AMEGO,  was selected through the NASA Astrophysics Research and Analysis (APRA) program to demonstrate the technology required for AMEGO. The ComPair project was initially funded in 2015 and underwent a high-energy beam test in 2022 (described in \cite{kirschner2024}), before its balloon flight from Fort Sumner, NM in 2023, which tested AMEGO's detector technology in a spacelike environment. 

In this paper, we will introduce the ComPair instrument and detail the detector subsystems in Section 2. An overview of the calibration campaign will be given in Section 3, including a description of the simulations performed. The analysis pipeline is described in Section 4, and the results of the calibration campaign will be presented in Section 5.


\section{The ComPair Instrument}
\label{sec2}

\begin{figure}[h]
  \centering
  \includegraphics[width=0.7\textwidth]{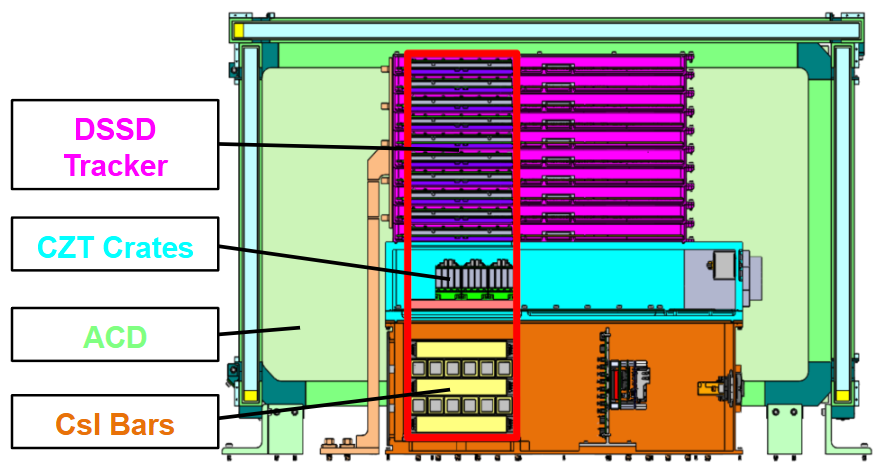}
  \caption{A CAD model of ComPair with the active area boxed in red. Each subsystem's detectors and electronics are integrated into aluminum enclosures stacked on top of each other, including the DSSD Tracker (magenta), the CZT Calorimeter (teal), and the CsI Calorimeter (orange). Surrounding the instrument stack is the ACD (green) that rejects charged particles.}
  \label{fig:ComPair_CAD}
\end{figure}

As seen in Fig.~\ref{fig:ComPair_CAD}, ComPair is composed of smaller versions of AMEGO's four subsystems. The double-sided silicon strip detector (DSSD) Tracker \cite{kirschner2024, Griffin:ComPair} is comprised of 10 single-detector layers stacked on top of each other. The cadmium zinc telluride (CZT) Calorimeter \cite{FrischGridCZT}, composed of 9 crates of 4$\times$4 arrays of CZT bars, is placed underneath the Tracker. The cesium iodide (CsI) Calorimeter, sitting below the Tracker and CZT Calorimeter, consists of 5 layers of six CsI bars where each layer is orthogonal to its neighboring layers in a hodoscopic arrangement \cite{Woolf_2018, Shy_2023}. Finally, an Anti-Coincidence Detector (ACD) surrounds the subsystem stack to reject the charged particle background \cite{metzler2024}. While the ComPair instrument is a new concept, it is important to note that it takes design heritage from previous MeV instruments with a DSSD tracker and CsI calorimeter, such as MEGA \cite{Bloser_2003} and TIGRE \cite{TIGRE}. The ComPair subsystems utilize modern advancements, with SiPM readout for the ACD and CsI calorimeter, in addition to low-noise ASICs for the DSSDs. ComPair's novel CZT calorimeter further enhances the instrument's capabilities in the Compton regime. 

\begin{figure}[h]
  \centering
  \includegraphics[width=0.3\textwidth]{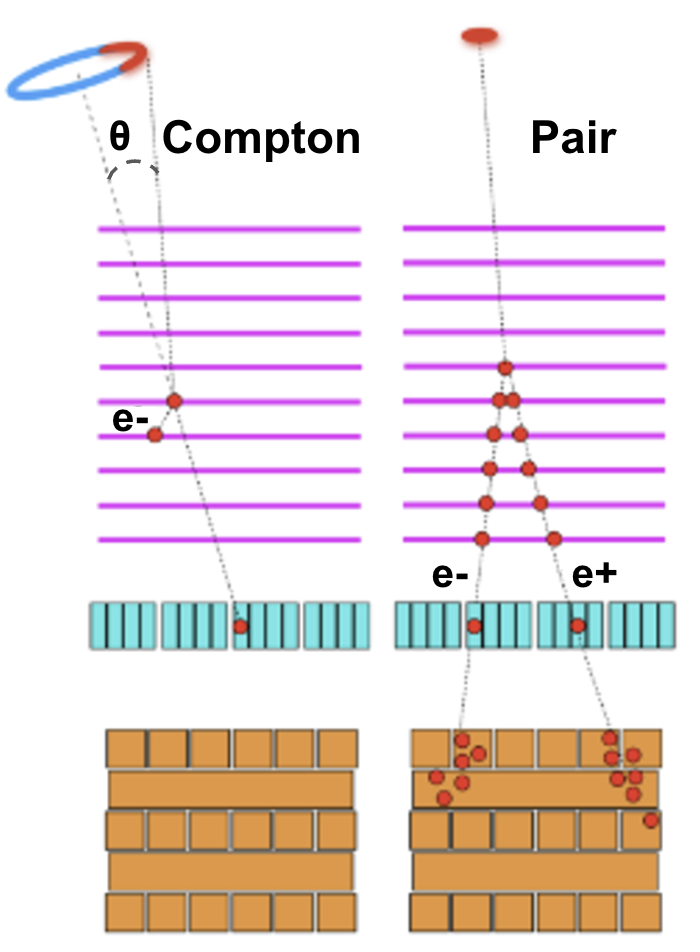}
  \caption{ComPair's design enables reconstruction of both Compton and pair events. Below $\sim$10 MeV, photons are more likely to Compton scatter in the Tracker (magenta), before the scattered photon gets absorbed in the CZT Calorimeter (teal). Above $\sim$10 MeV, photons will likely convert into an electron-positron pair in the Tracker, and the pair products are tracked through the CZT Calorimeter before being absorbed in the CsI Calorimeter (orange) \cite{Valverde_2023}.}
  \label{fig:ComPair_Modes}
\end{figure}
 
 Figure~\ref{fig:ComPair_Modes} details how ComPair detects both Compton scattering and pair production events. For Compton events, a photon can interact with at least one of the Tracker layers, scattering an electron in the process, before being absorbed by the CZT and/or CsI Calorimeter. The energy of the initial photon is found by summing the energies of each interaction. The positions of the photon interactions are used to recover the initial direction of the incident photon via the Compton equation:

\begin{equation}
\label{eq:Compton}
\theta = \cos^{-1} \left[ 1 - \left( \frac{1}{E_{\gamma'}} - \frac{1}{E_\gamma} \right) m_e c^2 \right]
,
\end{equation}
where $E_\gamma$ is the energy of the incident photon, $E_{\gamma'}$ is the energy of the scattered photon, $\theta$ is the scattering angle between the scattered photon and the initial direction, $m_e$ is the rest mass of an electron, and $c$ is the speed of light.

For pair production events, a photon interacts with a Tracker layer creating an electron and positron pair. This pair leaves 2 ionization tracks through the Tracker, before being absorbed in the calorimeters. Once again, the incident photon energy is found by summing the energies of each interaction. The trajectories of both the positron and electron are used to recover the initial direction via conservation of momentum.

\subsection{The Silicon Tracker}
\label{sec2.1}
The ComPair Tracker allows for the most accurate position resolution of the subsystems and is sensitive from $\sim$40 keV to $\sim$700 keV. The Tracker is composed of 10 layers of Micron Semiconductor double-sided silicon strip detectors (DSSDs). Each detector is $10 \times 10$~cm$^{2}$ and $500 ~\mu\text{m}$ thick with 192 orthogonally-oriented strips per side with a pitch of $510 ~\mu\text{m}$.  The orthogonal strips allow for the measurement of the X and Y position of interaction within the wafer. The Tracker has a position resolution of 300 $\mu$m FWHM and an energy resolution of 4\% FWHM at 122 keV \cite{kirschner2024}.

\begin{figure}[H]
\centering
\vspace{-.05cm}
    \includegraphics[width=0.45\textwidth]{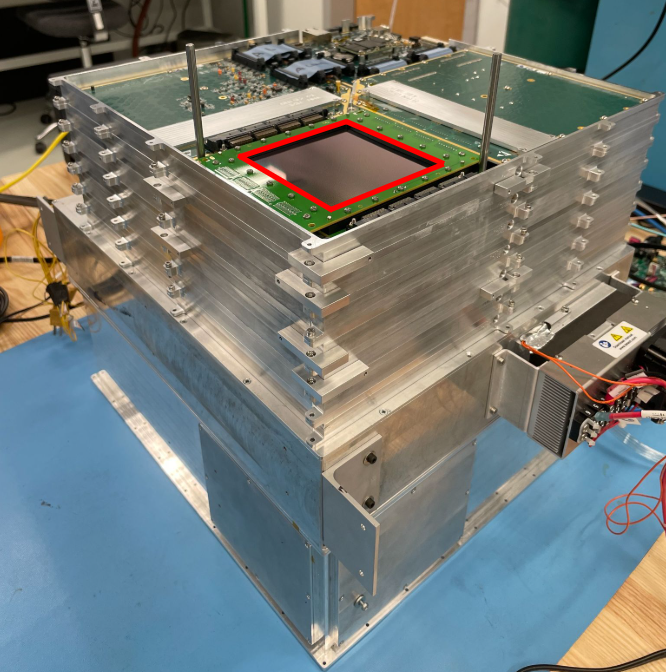}
\caption{Eight out of 10 layers of ComPair's silicon Tracker stacked on top of the CZT Calorimeter. The DSSD (boxed in red) takes up a quadrant of each layer, with the surrounding readout electronics taking up the other 3 quadrants.}
\label{fig:TKR_Stack}
\end{figure}

Figure~\ref{fig:TKR_Stack} shows 8 of ComPair's 10 Tracker layers during integration with the top layer open, exposing one of the DSSDs (boxed in red). The detectors are reverse biased with a positive voltage of 60V in order to fully deplete the 500 $\mu$m depth of the detector. The read-out, consisting of IDEAS VATA460.3 ASICs \cite{ASIC}, are AC coupled to the strips. The ASIC data is transmitted to a Xilinx Zynq 7020 field programmable gate array (FPGA) on each layer for processing. The detectors are mounted in a custom carrier board using elastomeric electrical contacts to supply the bias voltage. To see more details about the ComPair silicon Tracker, see \cite{kirschner2024}.

\subsection{The CZT Calorimeter}
\label{sec2.2}

\begin{figure}[h]
  \centering
  \begin{minipage}[t]{0.60\textwidth}
    \centering
    \includegraphics[width=\textwidth]{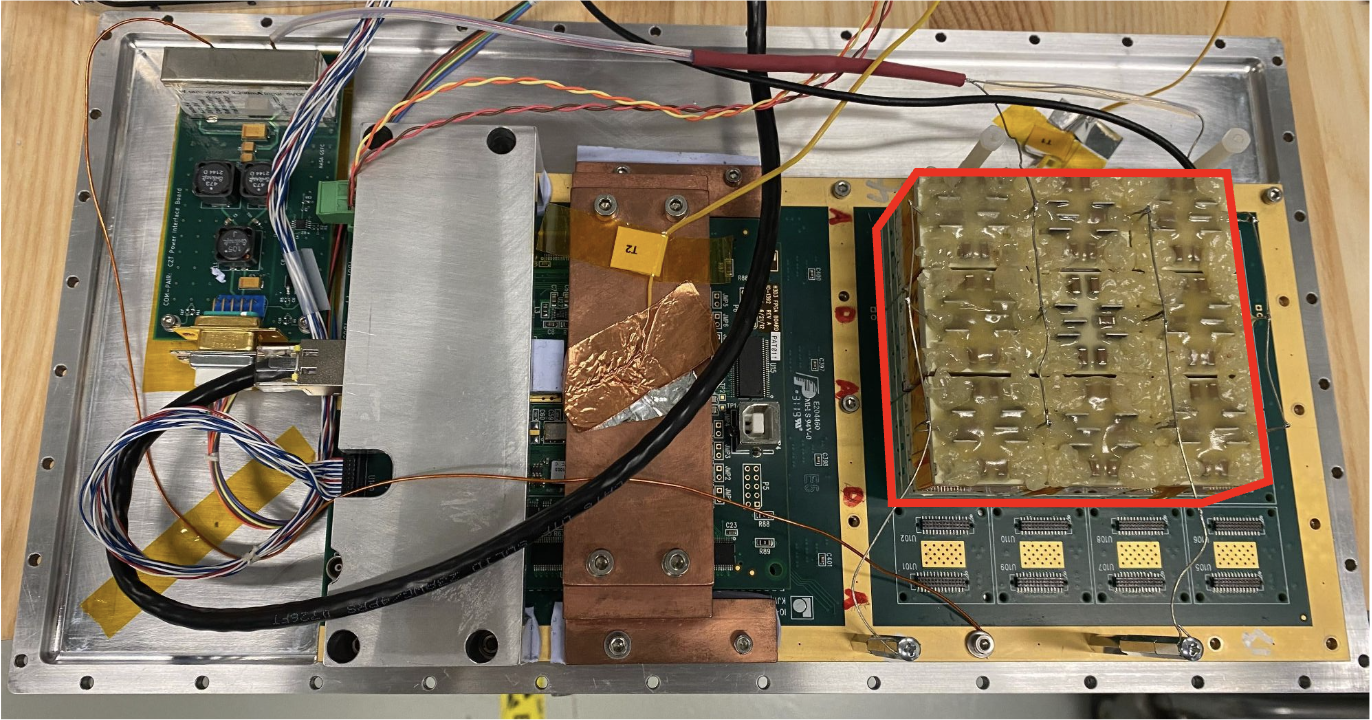}
    \caption{The inside of ComPair's CZT Calorimeter pressure vessel enclosure. The active detector area, composed of 9 crates, is boxed in red. The HV supply and FPGA are located on the left side of the enclosure.}
    \label{fig:CZT_Hardware}
  \end{minipage} \hfill
  \begin{minipage}[t]{0.36\textwidth}
    \centering
    \includegraphics[width=\textwidth]{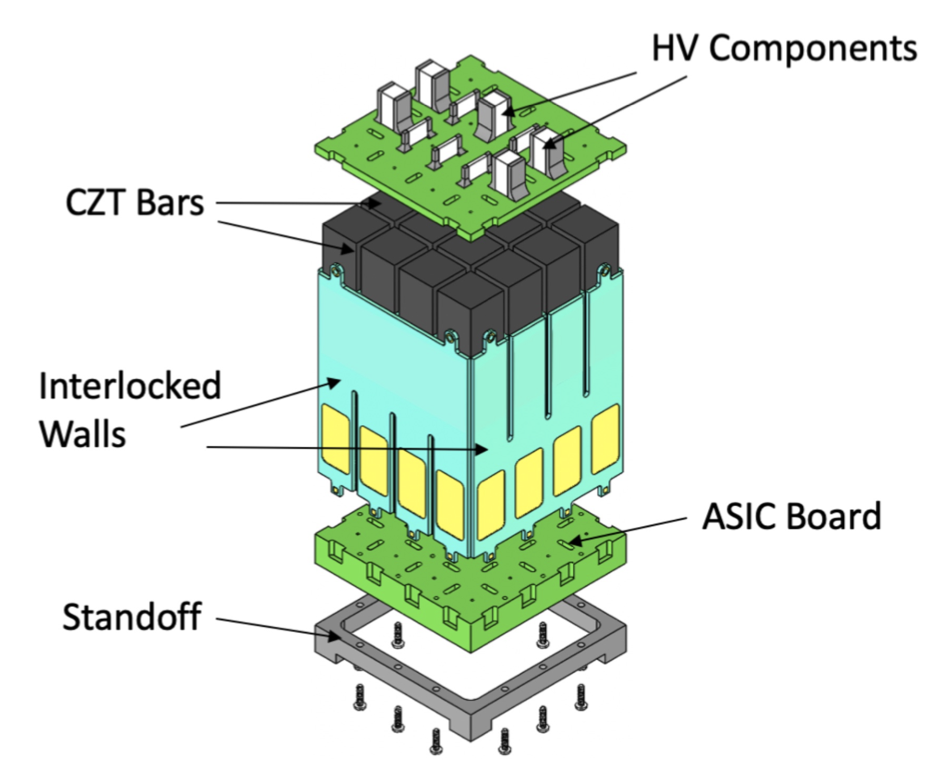}
    \caption{An exploded view of a CZT crate, showing the 4$\times$4 CZT bars \cite{AMEGO_Kierans}.}
    \label{fig:CZT_Exploded}
  \end{minipage}
\end{figure}

The CZT Calorimeter, shown in Fig.~\ref{fig:CZT_Hardware}, is composed of 9 crates, arranged in a 3 $\times$ 3 array nearly spanning the area subtended by the Si Tracker DSSDs. As seen in Fig.~\ref{fig:CZT_Exploded}, each crate contains a 4 $\times$ 4 array of 0.6 $\times$ 0.6 $\times$ 2 cm$^3$ CZT bars, where each bar has a gold contact on the top (cathode) and bottom (anode) connected to an external high-voltage source to bias the detectors and a low noise ASIC, respectively. Due to limitations in the ASIC readout, each 2 $\times$ 2 array of CZT bars share a single cathode. The custom analog ASICs were developed by Brookhaven National Laboratory (BNL) \cite{CZT_ASIC}. Four metal pads are placed around each CZT crystal to act as a virtual Frisch-grid \cite{BOLOTNIKOV_2024}. This allows for the evaluation of the X and Y coordinates of an interaction from the amplitude of the signals induced on the side electrodes. The Z coordinate of an interaction can be calculated from the relative cathode to anode pulse height ratio. 

The ComPair CZT Calorimeter sits in the middle of the detector stack, above the CsI Calorimeter, and below the Tracker. In order to mitigate the risk of high voltage breakdown at balloon flight altitude pressure, it was enclosed in a stand-alone pressure vessel. It has a dynamic range of 200 keV - 10 MeV, encapsulating the Compton-dominated regime. The CZT Calorimeter has an energy resolution of 2\% FWHM at 662 keV and a 3D position resolution of 0.2 cm FWHM. 

\subsection{The CsI Calorimeter}
\label{sec2.3}

\begin{figure}[H]
  \centering
  \begin{minipage}[t]{0.48\textwidth}
    \centering
    \includegraphics[width=\textwidth]{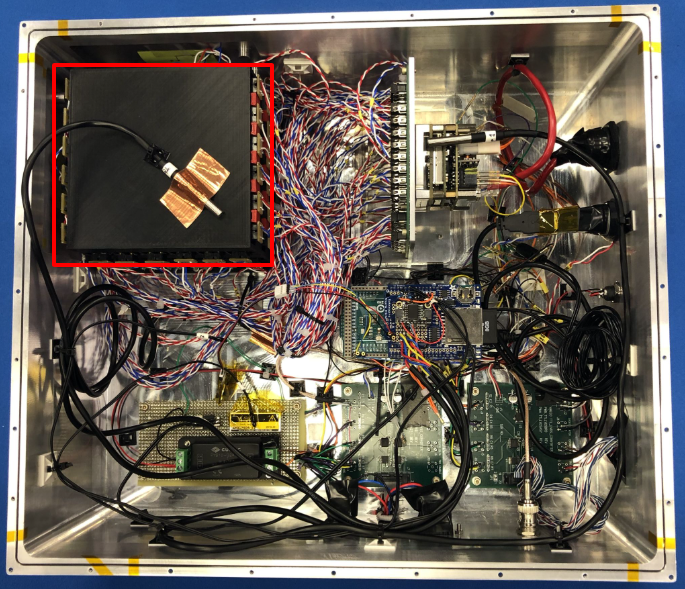}
    \caption{The inside of the ComPair CsI Calorimeter. The active detector area is boxed in red. The ROSSPAD SiPM readout is just above the hodoscope.}
    \label{fig:CsI_Hardware}
  \end{minipage} \hfill
  \begin{minipage}[t]{0.48\textwidth}
    \centering
    \includegraphics[width=\textwidth]{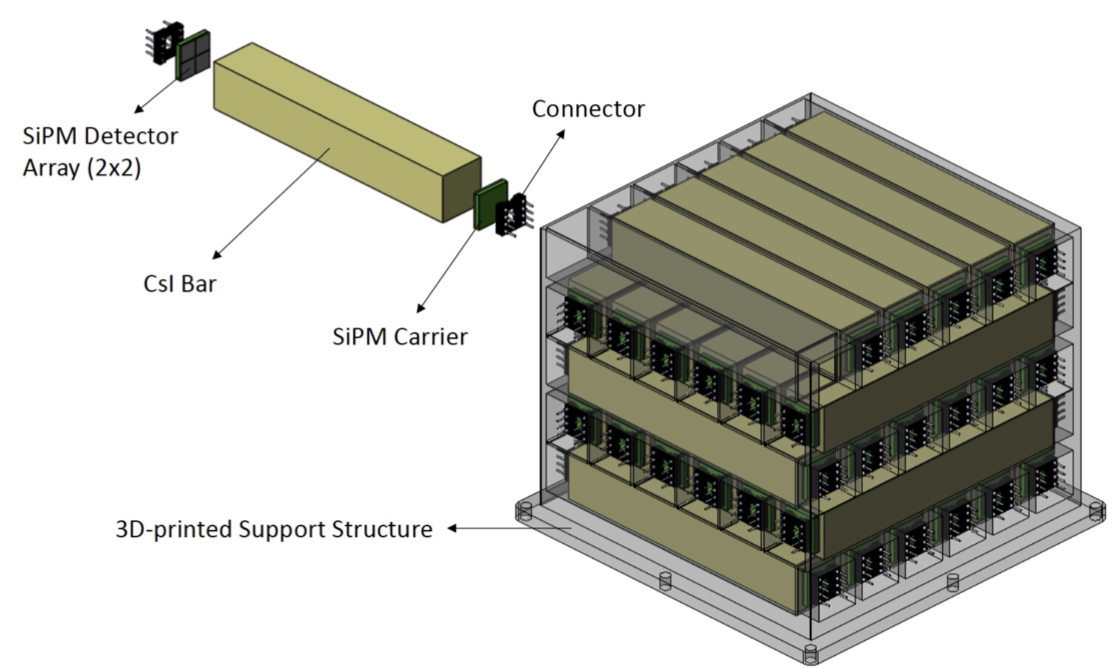}
    \caption{The design of the CsI detector, illustrating the hodoscopic arrangement of bars with the SiPMs mounted to each end \cite{AMEGO_Kierans}.}
    \label{fig:CsI_Design}
  \end{minipage}
\end{figure}

Figure~\ref{fig:CsI_Hardware} details the inside of ComPair's CsI Calorimeter enclosure, where the detector is boxed in red. The detector consists of 5 layers of six 1.7 $\times$ 1.7 $\times$ 10 cm$^3$ CsI bars arranged hodoscopically, where each layer is orthogonal to its neighboring layer, as seen in Fig.~\ref{fig:CsI_Design}. The bars are read out by ONSemi J-Series silicon photomultipliers (SiPMs) on each end, and the SiPMs are read out by an IDEAS ROSSPAD. The relative amplitude of the signals on either end of the bar is then used to calculate the depth of the interaction along the crystal (i.e., X and Y coordinates) \cite{Woolf_2018}. The Z position of an interaction is taken to be at the center of the bar.

The CsI Calorimeter sits at the bottom of ComPair's detector stack and has a dynamic range of 0.25 - 30 MeV per bar, allowing it to recover a photon's initial energy up to $\sim$100 MeV. The CsI Calorimeter has an energy resolution of 2.9\% FWHM at 1.27 MeV and a position resolution of along the bar of $\sim$1 cm FWHM. To see more details about the ComPair CsI Calorimeter and it's performance in the balloon flight, see \cite{Shy_2023, shy2024}.

\subsection{The Anti-Coincidence Detector (ACD)}
\label{sec2.4}

\begin{figure}[h]
\centering
    \includegraphics[width=0.45\textwidth]{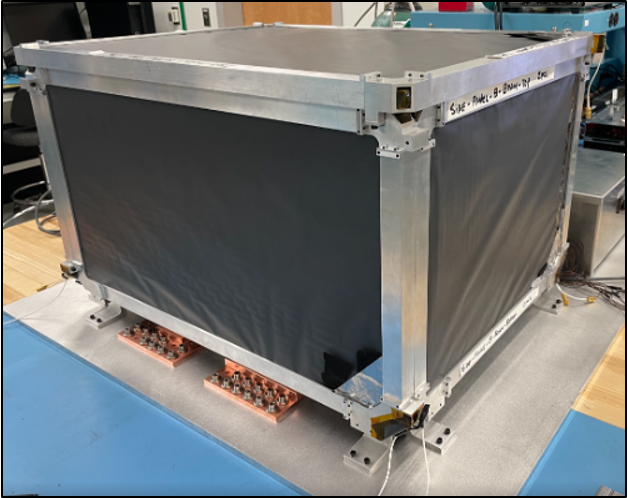}
\caption{ComPair's ACD fully surrounds the main instrument stack. Each plastic scintillator panel was wrapped in Tedlar (black wrapping) to be light-tight.}
\label{fig:ACD}
\end{figure}

Since ComPair's detectors depend on tracking charged particles generated from Compton or pair production interactions, the rejection of charged cosmic-ray background particles, such as protons and muons, is critical. This necessitates the need for the ACD, which encapsulates the rest of the subsystems. Figure.~\ref{fig:ACD} shows the integrated ACD, comprised of plastic scintillators (Eljen EJ 208 \cite{EJ208}) with wavelength shifting bars (Eljen EJ 280 \cite{EJ280}), read out by SiPMs \cite{AMEGO}. The read out electronics are identical to the readout electronics in the CsI Calorimeter.

Any event detected by the main ComPair instrument stack that has a simultaneous interaction in the ACD is vetoed, which reduces the charged particle background. To see more details about the ComPair ACD and it's performance in balloon flight, see \cite{metzler2024}.

\subsection{Data Acquisition}
\label{sec2.5}

 Each of the 10 Tracker layers, and the 3 other subsystems are connected to a Trigger Module (TM), which utilizes an FPGA development board (ZC706, Xilinx) to align events in each subsystem \cite{Makoto_2020}. The TM checks for coincidence between Tracker layers and other detector systems, which drastically suppresses noise in the data. Different trigger conditions are set based on the testing and operating configuration. Typical operation requires either two triggers in the Tracker, one on each side of a single detector layer, or coincidence between any two of the Tracker, CZT Calorimeter, and CsI Calorimeter. If either condition is met, an event ID is sent to each system for offline alignment. This was the configuration used for the full-system calibrations. Additionally, each subsystem can operate independently as a stand-alone system for subsystem specific calibrations \cite{kirschner2024, Shy_2023}.

\section{Calibration Campaign}
\label{sec3}

Calibrations play a key role in assessing and quantifying an instrument's performance. This allows for comparisons to be made with past and future missions. Since ComPair is the first prototype of AMEGO, and its successor, ComPair-2 \cite{Caputo_2024}, is already being developed, it is important to demonstrate measurable improvements with each iteration. Furthermore, the scientific returns of ComPair depend on the accuracy of the direction and energy measurements of the incident gamma-ray photons as well as the efficiency of the detectors. To quantify these factors, the Compton angular resolution, energy resolution, and effective area were measured prior to the balloon flight in the summer of 2023. 

\subsection{Setup and Methods}
To properly characterize the instrument's performance, laboratory radioactive sources (sealed Type D disk gamma-ray sources from Eckert \& Ziegler\footnote{\texttt{https://www.ezag.com/wp-content/uploads/2023/08/Gamma\_Standards\_D-Type.pdf}}) with a large range of line energies were used, listed in Table~\ref{tab:Calibration_sources}. These sources had line energies spanning from $\sim500$ keV - $\sim1300$ keV which adequately covers the Compton regime.  Note, the high thresholds of the CZT and CsI calorimeters limited the lower energy range in the Compton regime.

\begin{table}[h]
\centering
\resizebox{0.65\textwidth}{!}{%
\begin{tabular}{|c|c|c|c|c|}
\hline
\textbf{Isotope} & \textbf{Half-life} & \textbf{Line Energies (keV)} & \textbf{Transition Probability} & \textbf{Activity (Ci)} \\ \hline
$^{137}$Cs                  & 30.17 years                  & 661.7 & 0.85 & 6.39E-04                 \\ \hline
\multirow{2}{*}{$^{22}$Na}  & \multirow{2}{*}{2.6 years}    & 511   & 1.798 & \multirow{2}{*}{4.64E-04} \\ \cline{3-4}
                        &                               & 1274  & 0.999 &                          \\ \hline
\multirow{2}{*}{$^{60}$Co}  & \multirow{2}{*}{5.27 years}   & 1173.2 & 0.998 & \multirow{2}{*}{5.54E-04} \\ \cline{3-4}
                        &                               & 1332.5 & 0.999 &                          \\ \hline
\end{tabular}
}
\caption{Radioactive sources used for the full instrument calibration campaign.}
\label{tab:Calibration_sources}
\end{table}

These sources were used to quantify instrument performance as a function of energy and off-axis angle. As shown in Table~\ref{tab:Calibration_source_runs}, data was collected using each source directly above the detector. Additionally, $^{137}$Cs was used to quantify the instrument's off-axis response.

\begin{table}[h]
\centering
\resizebox{0.45\textwidth}{!}{%
\begin{tabular}{|c|c|c|c|}
\hline
\textbf{Isotope} & \textbf{Off-axis Angle} & \textbf{Height (cm)} & \textbf{Run length} \\ \hline
$^{137}$Cs       & 0$^\circ$   &    169.3 $\pm$ 0.1    & 12 hours  \\ \hline
$^{22}$Na      & 0$^\circ$  &    339.0 $\pm$ 5.0    & 12 hours  \\ \hline
$^{60}$Co      & 0$^\circ$  &    339.0 $\pm$ 5.0     & 10 hours  \\ \hline
$^{137}$Cs       & 10$^\circ$  &    169.3 $\pm$ 0.1  & 1.5 hours  \\ \hline
$^{137}$Cs     & 20$^\circ$  &    169.3 $\pm$ 0.1  & 1.5 hours  \\ \hline
$^{137}$Cs     & 30$^\circ$   &    169.3 $\pm$ 0.1  & 1.5 hours  \\ \hline
$^{137}$Cs      & 40$^\circ$  &    169.3 $\pm$ 0.1   & 1.5 hours  \\ \hline
$^{137}$Cs     & 50$^\circ$  &    169.3 $\pm$ 0.1     & 2 hours  \\ \hline
\end{tabular}
}
\caption{Radioactive source runs performed during the calibration campaign. The activities of the $^{22}$Na and $^{60}$Co sources were too high for them to be placed in the source structure, so instead these sources were placed in the rafters of the lab. The height of each source is measured from the bottom layer of the Tracker.}
\label{tab:Calibration_source_runs}
\end{table}

\begin{figure}[h]
\centering
    \includegraphics[width=0.45\textwidth]{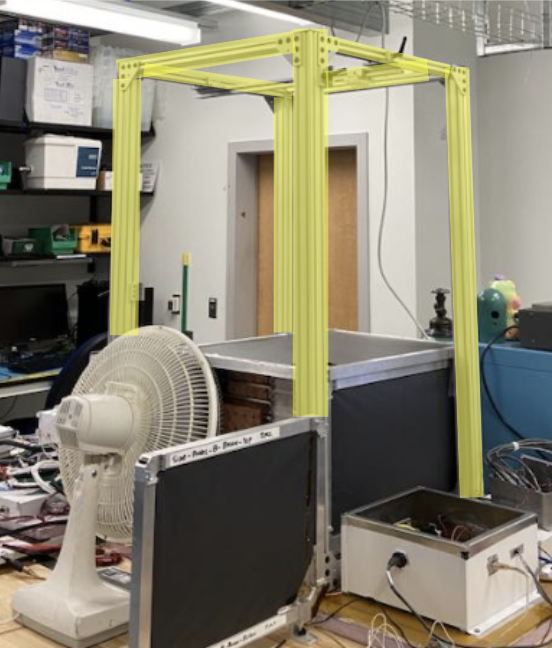}
\caption{Calibration structure (highlighted) placed around the full ComPair instrument. This structure was used to hold radioactive sources in precise locations during calibrations. The main instrument stack is surrounded by the ACD (in black). One of the ACD panels is open, allowing a fan to cool the instrument to prevent overheating during operation. The ACD electronics box (white box) sits beside the ACD.}
\label{fig:Calibration_structure}
\end{figure}

To ensure accurate measurements, the radioactivate sources had to be placed in precise locations. This was achieved with a calibration structure that was built around the instrument, as shown in Fig.~\ref{fig:Calibration_structure}. With the help of a plumb bob, the source could be placed directly over the center of the detector. However, the activities of the $^{22}$Na and $^{60}$Co sources were too high when placed in the source structure. Instead, these sources were placed up in the rafters of the lab. The plumb bob still allowed for the sources to be placed directly over the detector. However, the height measurement was less precise, which is why the error is larger for the $^{22}$Na and $^{60}$Co compared to the $^{137}$Cs source (which was used in the calibration structure). 

\subsection{Simulations}

\begin{figure}[H]
  \centering
  \includegraphics[width=0.45\textwidth]{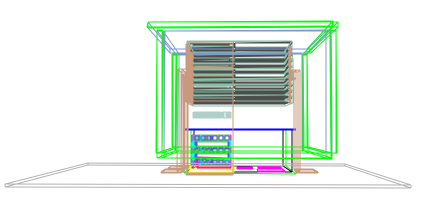}
  \caption{Mass model of ComPair created using MEGAlib's Geomega tool. The ACD is in green, and the Tracker with the aluminum housing takes up the top half of the instrument stack. The CZT bars are shown in teal, and the CsI Calorimeter with its aluminum box is at the bottom. All are mounted to a bottom aluminum base plate that was used for calibrations and during the balloon flight.}
  \label{fig:Geomega_MM}
\end{figure}

The ComPair calibration simulations make use of the software package MEGAlib \cite{Zoglauer_2005}. The MEGAlib tool Geomega, based on Geant4 (v10.02), was used to construct the mass model of the instrument, shown in Fig.~\ref{fig:Geomega_MM}. MEGAlib's Cosima was used to simulate the laboratory sources and track the radiation-matter interactions within the instrument (using Geant4 physics list G4EmLivermorePolarizedPhysics).

The Cosima output is the idealized case of interactions in the detectors. Therefore, in order to match the real data, a custom python-based detector effects engine (DEE) was developed to apply the performance of the detectors to the simulation output. The DEE takes in the idealized positions and energies from Cosima, applies an inverse calibration for each subsystem which converts energy to pulse height and x, y, z position to strip ID/detector ID/layer ID. Additionally, it adds realistic noise and thresholds, and converts the output to match the format of the raw detector data. After the simulations are run through the DEE, they are then run through the same forward calibration pipelines as the real data.

Simulations were used to estimate the data run lengths, as well as to verify calibration results. Previous work showed that 10,000 events in a photopeak was enough statistics to fit the data with a Gaussian, therefore simulations (without the DEE) were used to obtain rough estimates of the time needed to obtain this for each calibration run. For the data runs taken with the source directly above the detectors however, this time was significantly lengthened to ensure plenty of statistics for not only the angular resolution measurements (ARMs), but for the effective area measurements as well. 

After the calibration data was taken, more in-depth simulations were used to compare with the real data. One of the goals of calibrations is to benchmark simulations, which requires iterative adjustments of the DEE to better match the lab measurements. While this is a time consuming process, the initial comparison between the simulations and calibration data has been promising, and will be discussed in Section 5.

\section{Processing of Data}

Each subsystem has its own custom Python pipeline that the subsystem-specific data is run through before combining with data from the other subsystems. These pipelines convert the raw data, in analog-to-digital converter (ADC) values to energy, and the relative signals on detector channels to positions (X, Y, Z). Once each subsystem has converted the hits into energies and positions, the events are combined using Event IDs from the TM. These subsystem pipelines are described below.

\subsection{Tracker Pipeline}

When a charge is induced in a silicon detector strip, it is denoted as a strip hit. These strip hits are digitized by ASICs and the first step in the Tracker pipeline is to convert the raw data in ADC values measured on each detector strip to energy in keV, using an energy calibration file. The energy calibration is generated using radioactive sources with known decay emission lines. Each strip records data in ADC units. The photopeaks of these sources are then fit with a Gaussian, and the centroid ADC value is mapped to the energy of the source. Four radioactive sources in total are used per strip, and a linear relation is used to describe the relation between ADC values and energy. 

\begin{figure}[h]
\centering
    \includegraphics[width=0.45\textwidth]{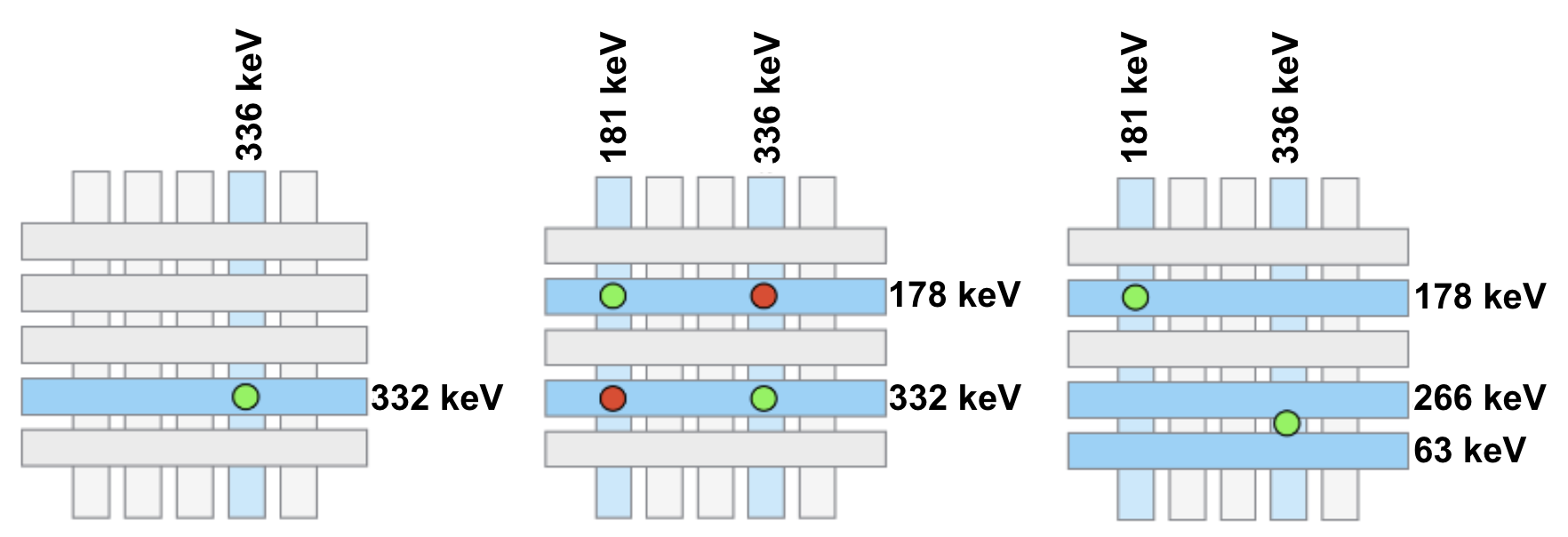}
\caption{(Left) For single interactions in the silicon, the position is found by taking the intersection of the strips, shown in green. (Center) For multiple interactions, multiple solutions exist, shown in red and green. In order to resolve the positions of the interactions, the strips that are closest in energy are paired together, thus the green circles are used in this case. (Right) When charge is shared between two adjacent strips, the position and energy must be resolved first before resolving multiple interactions. This is done by summing the energies of neighboring interactions and using a weighted average to obtain the position of the hit. Figures from \cite{Sleator_2019}.}
\label{fig:strip_pairing}
\end{figure}

The next step in the Tracker pipeline processes the data through a strip pairing algorithm that resolves each interaction, detailed below. In the simplest case of a single interaction, where a particle generates a strip hit on either side of the detector, the position is simply the intersection of the two strips. For instances of charge-sharing, where two neighboring strips share the energy from an interaction, the energy of the neighboring strip hits is summed and the position is taken as a weighted average between the two strips. This is done before resolving multiple interactions to more accurately match the energies. For multiple coincident interactions in a detector, where there are multiple solutions to the positions of the interactions, X and Y strips are paired together based on the measured energy (Fig.~\ref{fig:strip_pairing}). 

\subsection{CZT Calorimeter Pipeline}

For the CZT Calorimeter, when energy is deposited in a CZT bar, signals are induced on 6 electrodes: the anode, cathode, and 4 charge-sensitive pads placed on each side of the CZT bar. The anode signal corresponds most directly to the energy deposited and can be corrected based on positional information from the other 5 channels. The Z position of the interaction is obtained from the cathode-to-anode (C/A) ratio, and the X and Y positions are obtained using a center of mass calculation:
\begin{equation}
    X = \frac{Q_{X_1}}{Q_{X_1} + Q_{X_2}}, ~~Y = \frac{Q_{Y_1}}{Q_{Y_1} + Q_{Y_2}}
\end{equation}
Where $Q_{X_1}, Q_{X_2}, Q_{Y_1},$ and $Q_{Y_2}$ are the amplitudes of the signals on the 4 side pads. 

The CZT energy calibration utilizes data from the 662 keV line from $^{137}$Cs. The first step is to determine the pedestal location for the anode channel, which is done by fitting a Gaussian to the pedestal peak of the raw ADC spectrum. The pedestal value is treated as the 0 keV point for the energy calibration of the ADC spectra. Next, a depth-of-interaction correction is applied to each CZT measured anode ADC value to account for a loss of collected charge further from the bottom of the bar. The anode ADC pulse height corresponding to the 662 keV photopeak is recorded as a function of the C/A ratio, which is a proxy for depth of interaction in the CZT bar. A correction for each measured C/A ratio flattens this distribution such that the recorded anode pulse height is equivalent to the measured charge when the interaction occurs at the bottom of the bar closest to the anode. Finally a linear energy calibration is applied converting the depth-corrected ADC value to keV using the 662 keV photopeak.

Finally, two cuts are made to the CZT data. The first cut selects events for which all 4 pad signals are above a predetermined threshold. The second cut removes hits that are reconstructed to be at the top of the CZT bar. Since the depth of interaction comes from the C/A ratio, the range of values representing the top and bottom of the CZT bar is identified from the histogram of C/A values. Events that fall outside of this range get stacked at the top of the bar and therefore are removed. More information on calibrating Virtual Frisch Grid CZT detectors can be found in \cite{BOLOTNIKOV_2024}. Additionally, a more detailed description of ComPair's CZT calibration pipeline and performance is in preparation.

\subsection{CsI Calorimeter Pipeline}

When energy is deposited in the CsI bars, SiPMs read out the scintillation signal, $Q_{left}$ and $Q_{right}$, at each end of the CsI bars. The depth of interaction (DOI) in the bar is then found by,
\begin{equation}
    \mathrm{DOI} = \frac{Q_{left} - Q_{right}}{Q_{left} + Q_{right}},
\end{equation}
while the energy deposited is found with,
\begin{equation}
    \mathrm{E}_{ADC} = \sqrt{Q_{left} \times Q_{right}}.
\end{equation}

Position calibrations are then applied to the depth of interaction, and energy calibrations are applied to the E$_{ADC}$ as described in \cite{Shy_2023}. A GPS disciplined pulse per second (PPS) is used to correct the on board clock. The PPS in conjunction with the trigger acknowledge signal from the TM is used to align events in time to declare coincidence events with other instruments. However, this is only needed for the CsI Calorimeter and the ACD since Event IDs are written directly into the data for the other subsystems.

\subsection{ACD Pipeline}

The ACD uses SiPMs to read out the wavelength shifting bars.  When any individual SiPM array records an increase in signal above the trigger threshold ($\sim$3 MeV), all 10 SiPMs are read out and recorded. Additionally, the ACD receives a trigger acknowledge (TrigAck) signal from the TM and a PPS signal. This is the same read-out used by the CsI Calorimeter. The first part of the ACD pipelines matches the PPS signals in the ACD and TM, then assigns the respective Event ID to the TrigAck. Only events that occurred within 20$\mu$s of a TrigAck are kept, and are assigned the respective Event ID. Then, the energy calibration is applied to each hit.

The energy calibration file comes from a calibration run in which all 5 ACD panels are stacked on top of each other, and coincidence between all 5 panels is enforced to select transversing muons. This coincidence requirement significantly suppresses the noise pedestal in the spectra. These spectra are then compared to muon simulations performed in Cosima (see section 3.2 for more details). The real data in ADC-space and the simulated data in energy-space are both fit with Landau distributions as expected from minimum ionizing particles. From these fits, ADC-space is mapped to energy-space using two points: $\mu - \eta$ and $\mu$, where $\mu$ is the location of the peak and $\eta$ is the width of the distribution. See \cite{metzler2024} for more details on the ACD and its pipeline.

\subsection{Full Instrument Pipeline}
After the individual subsystem data is calibrated, such that they have positional and energy information for each event, this data is then combined on an event-by-event level to be read by Revan, a MEGAlib tool that performs event reconstruction \cite{Zoglauer_2005, ZOGLAUER_2006}.  Revan categorizes and reconstructs each event into one of four categories: single-site event, Compton scatter event, pair production event, or high-energy charged particle event (such as a muon). In this case, a single-site event refers to an event that interacts only once in the instrument. This includes photo absorption events and partially absorbed Compton events. Revan's specialization lies within Compton event reconstruction, where it contains several algorithms to determine the Compton interaction sequence of each event. After event reconstruction, the data is analyzed in Mimrec, MEGAlib's high-level analysis tool \cite{Zoglauer_2011_Mimrec}. Mimrec offers options to perform event selections, view spectra and ARM distributions, reconstruct images, and more. For each calibration run, Mimrec was used to generate the full instrument spectrum (which includes single-site photoabsorption events) and the Compton spectrum, in addition to the ARM distribution.

\section{Calibration Results}

\subsection{Energy Resolution}

\begin{figure}[H]
  \centering
  \begin{minipage}[t]{0.48\textwidth}
    \centering
    \includegraphics[width=\textwidth]{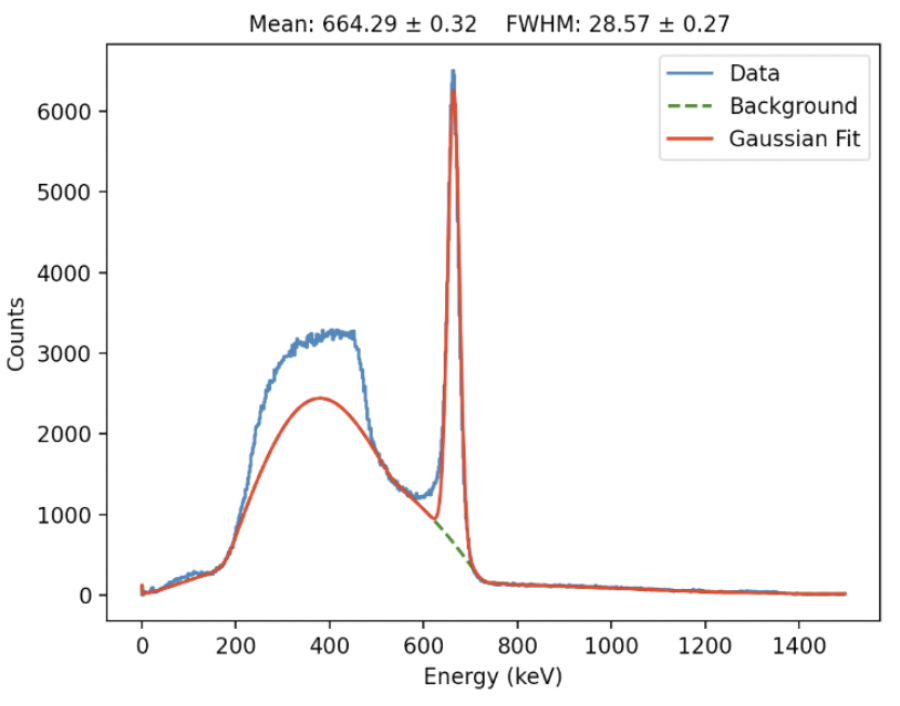}
    \caption{Full instrument spectrum of $^{137}$Cs, including both Compton and single-site events. A clear photopeak can be observed at $\sim$662 keV, along with the corresponding Compton edge at 480 keV. The background is displayed as a dotted green line. A Gaussian function was fit to the background-subtracted photopeak. The red line shows the Gaussian fit after adding back the background (green). The Compton edge was not included in the fit, which is why the data rises above the fit between 200 and 500 keV.}
    \label{fig:Cs137_Spectra}
  \end{minipage} \hfill
  \begin{minipage}[t]{0.48\textwidth}
    \centering
    \includegraphics[width=\textwidth]{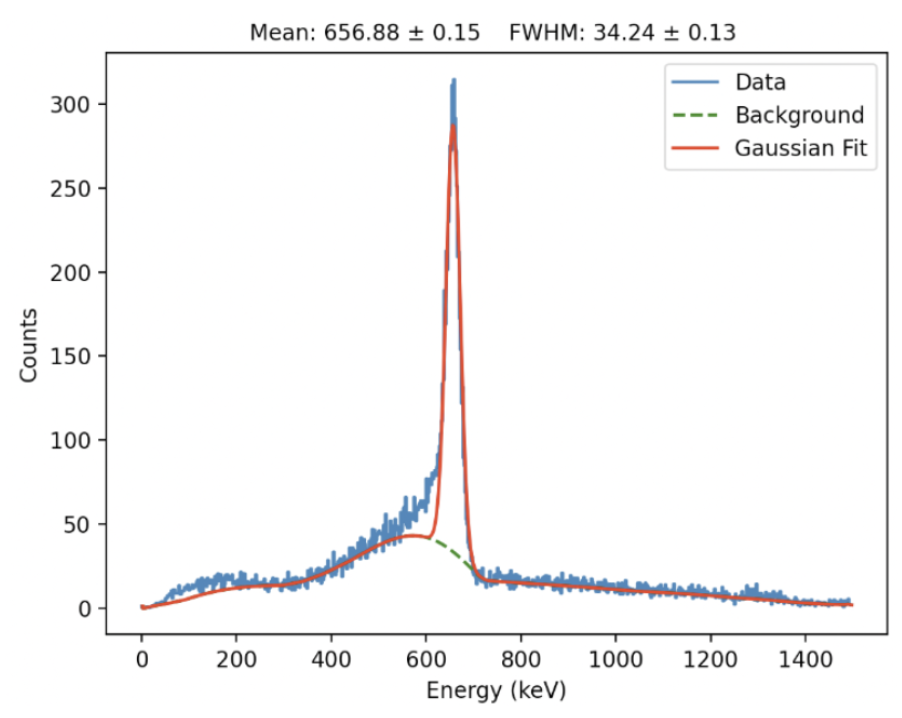}
    \caption{The full instrument Compton spectrum of $^{137}$Cs. Even after excluding single-site events, the Revan-identified Compton events still display a strong peak at $\sim$662 keV. Once again, the background was fit with ROOT (green dotted line) and the peak was fit with a Gaussian (red).}
    \label{fig:Cs137_Spectra_Compton}
  \end{minipage}
\end{figure}

The energy resolution calibration measured the energy response of the combined detectors. The energy response in the ideal case would be represented by a delta function at the radioactive source's energy. However, due to photoelectron statistics and inherent detector effects, the emission line is broadened into a photopeak, which can be fit with a Gaussian. After generating the full instrument spectra (including both Compton and single-site events) and the Compton-only spectra using Mimrec, the underlying Compton continuum in both cases was fit using ROOT’s built-in background fitting function. The resulting data was then extracted for further analysis in Python. The background-subtracted photopeaks were each with fit a Gaussian using the curve fit function from the SciPy python library, and the FWHM of the Gaussian fits were recorded. It is important to note that the Compton edge is not fully modeled in this fit function; however, as the goal is to find the underlying continuum to isolate the counts in the photopeak, the built-in background fitting function is sufficient. Figures~\ref{fig:Cs137_Spectra} and~\ref{fig:Cs137_Spectra_Compton} show the results of these fits after adding back the underlying continuum. As seen in Fig.~\ref{fig:Cs137_Spectra}, the energy resolution of ComPair at 662 keV is 28.57 keV $\pm$ 0.3 keV, or 4\% FWHM/E. 

\begin{figure}[h]
  \centering
  \begin{minipage}[t]{0.48\textwidth}
    \centering
    \includegraphics[width=\textwidth]{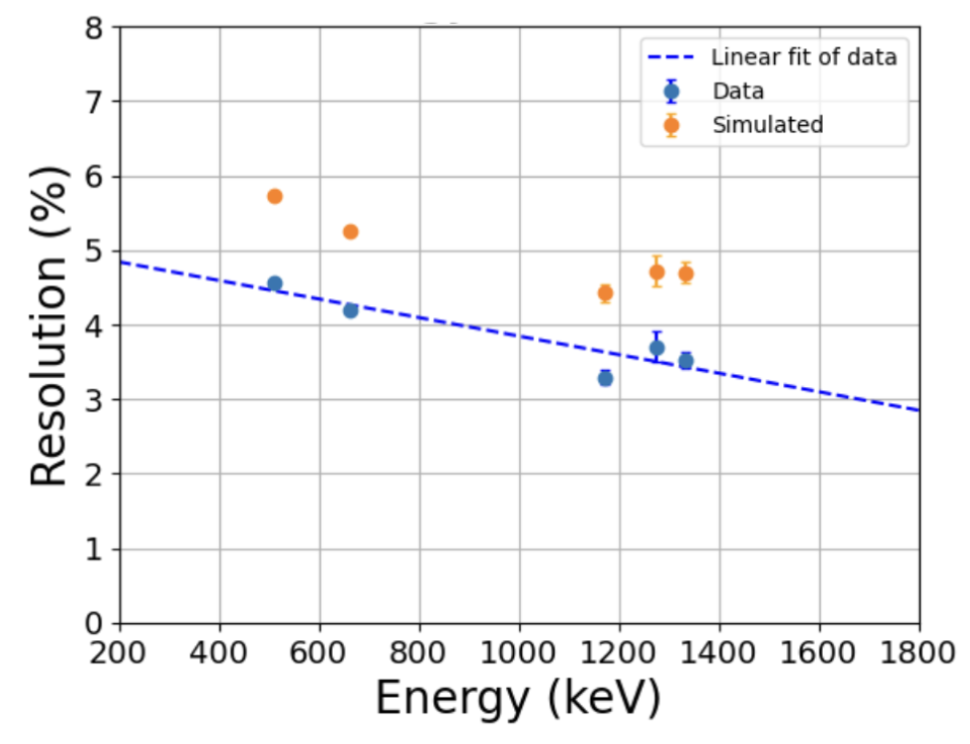}
    \caption{The measured full instrument energy resolution as a function of energy (blue) versus the simulated energy resolution (orange). The measured data was fit with a linear function to understand the trend.}
    \label{fig:ComPair_Energy_Resolution}
  \end{minipage} \hfill
  \begin{minipage}[t]{0.48\textwidth}
    \centering
    \includegraphics[width=\textwidth]{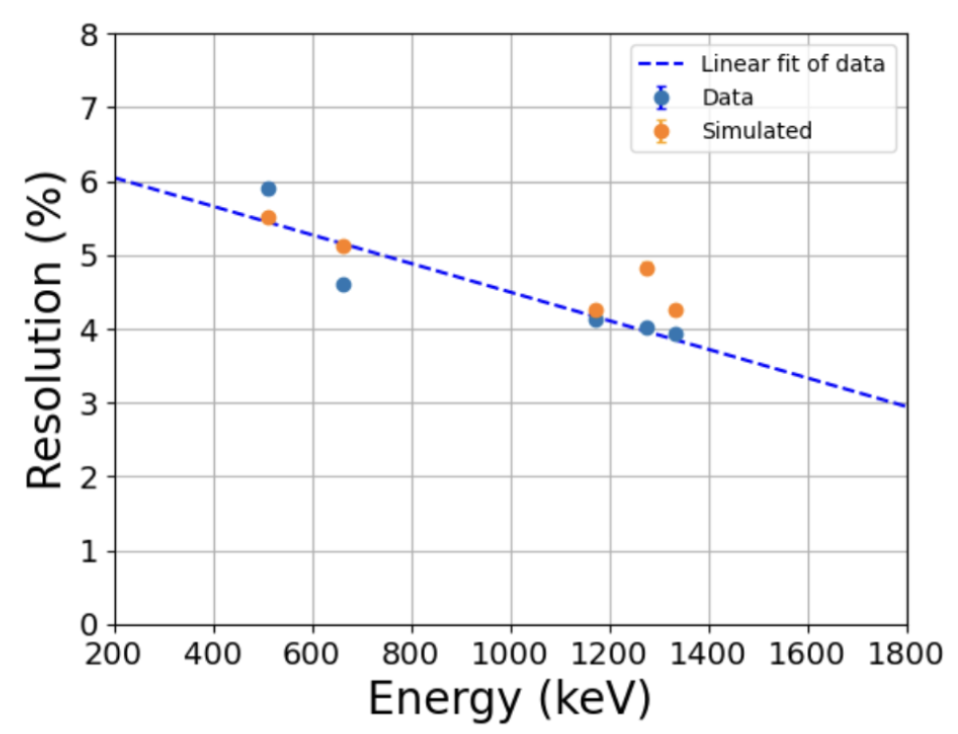}
    \caption{Linear fit of the measured Compton energy resolution as a function of energy (blue) versus the simulated Compton energy resolution (orange).}
    \label{fig:ComPair_Energy_Resolution_Compton}
  \end{minipage}
\end{figure}

Figures~\ref{fig:ComPair_Energy_Resolution} and \ref{fig:ComPair_Energy_Resolution_Compton} show the full instrument energy resolution and the Compton energy resolution, respectively, as a function of energy, where the energy resolution FWHM in keV is divided by the energy of the photopeak for each calibration source. The error bars were calculated directly from the fitting routine and represent the uncertainties in the fitted parameters. Note that the error bars for some data points are too small to be visible on the plot. The full instrument energy resolution can use photoabsorption events from individual subsystems whereas the Compton energy resolution requires multiple interactions, therefore the Compton energy resolution is expected to be worse, since the energy resolution of each interaction is added in quadrature. To capture the general trend in resolution performance across the measured energy range, a linear fit was performed on both data sets.

In Cosima, we simulated the source with its real flux, using a stop criteria of 1 hour to get sufficient statistics. As detailed in Section 3.2, these simulations were sent through the DEE before being run through the subsystem calibration pipelines and Revan, therefore the performance should be similar to the measured data. The same procedure was used to obtain energy resolution measurements from the simulations, which can be compared to the data. See Appendices A and B for the full instrument and Compton-only spectra, respectively, for all calibration sources.

The full energy resolution and the Compton energy resolution were found to be comparable with simulations. However, benchmarking simulations is an iterative process that requires adjusting the DEE to better match the real data. Current simulations match the shape of the full system energy resolution, with a small systematic overestimation. This overestimation occurs in order for the Compton energy resolution to match. This discrepancy is not fully understood, however, it is important to note that the DEE development is still a work in progress. 

\begin{figure}[h]
  \centering
  \includegraphics[width=0.45\textwidth]{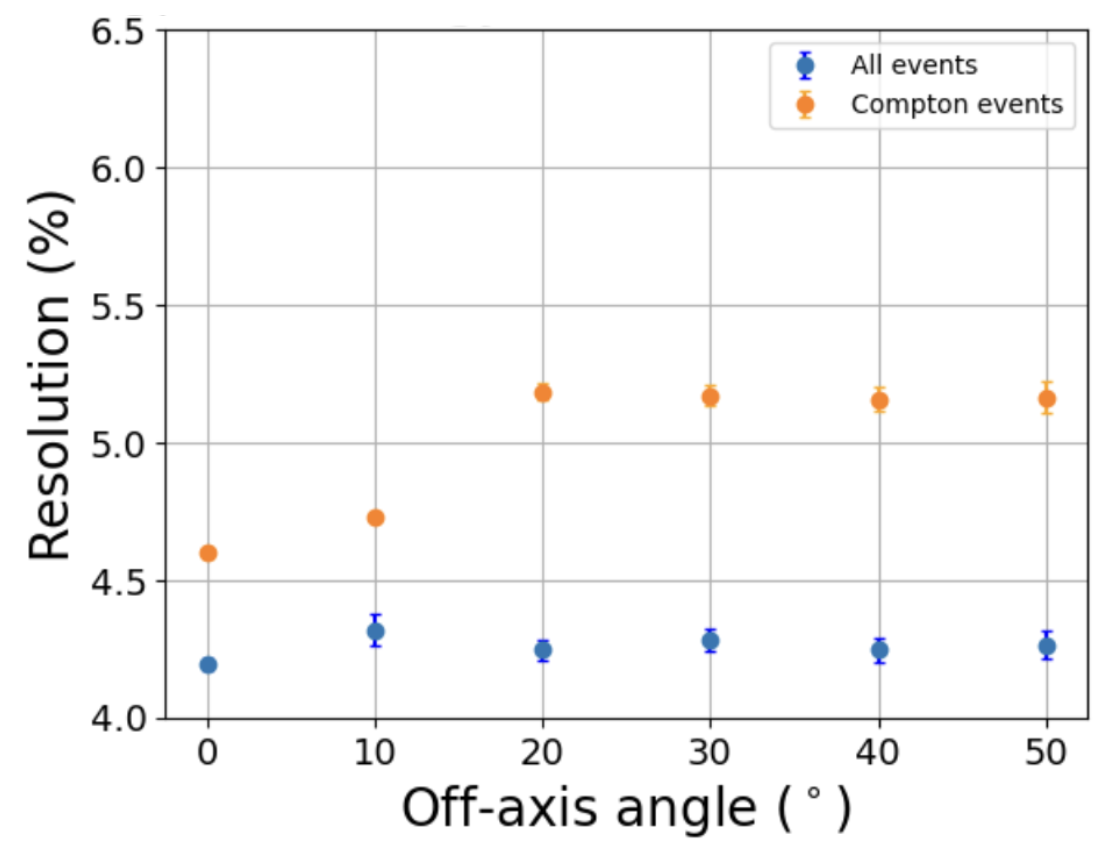}
  \caption{Measured energy resolution as a function of off-axis angle at 662 keV for all events (blue) and Compton events (orange). The trend in the Compton energy resolution is not well understood.}
  \label{fig:off-axis_EnergyRes}
\end{figure}

Using the same procedure, the full energy resolution and Compton energy resolution were also found as a function of off-axis angle, as shown in Fig.~\ref{fig:off-axis_EnergyRes}, for the $^{137}$Cs 662 keV line. Although the full instrument energy resolution remained stable across all off-axis angles, the Compton energy resolution degraded with increasing angle up to 20$^\circ$, after which it plateaued. This trend is currently being investigated to determine whether it is statistically significant or an unidentified systematic bias. 

\subsection{Angular Resolution}

\begin{figure}[h]
  \centering
  \includegraphics[width=0.45\textwidth]{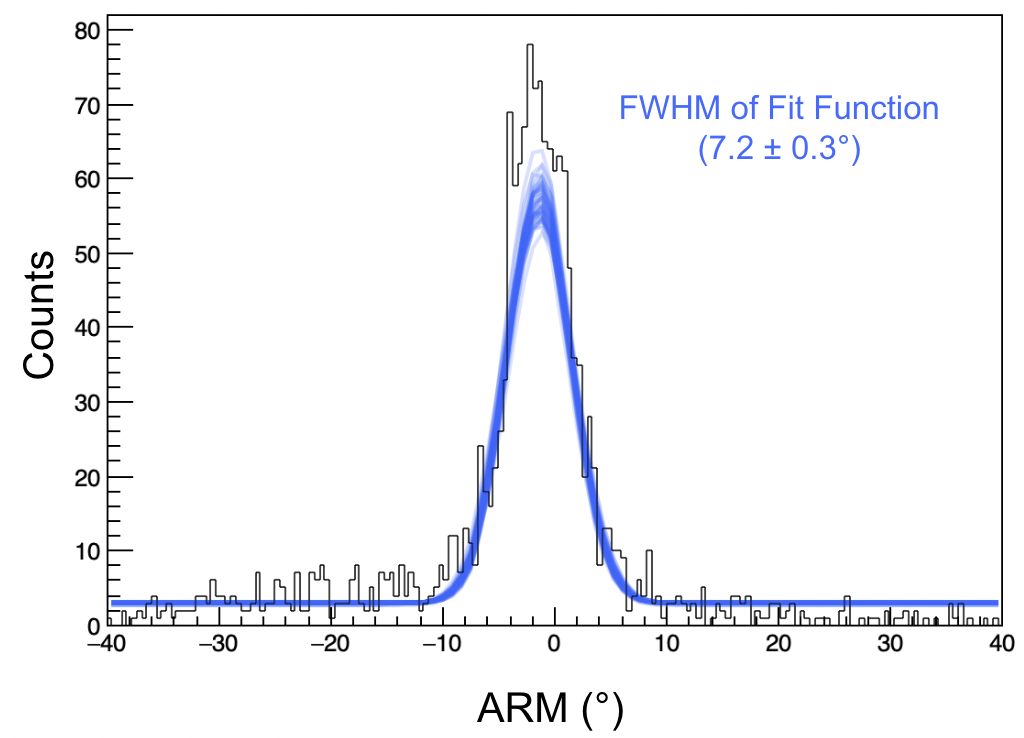}
  \caption{Histogram of ARM values from an on-axis $^{137}$Cs source. For this dataset, the first interaction was required to be in the Tracker, followed by an interaction in the CZT Calorimeter. Additionally, a cut requiring the Compton scattering angle to be less than 60$^\circ$ was made. The uncertainty in the FWHM of the fit was estimated by bootstrapping the data and performing 50 repeated fits, shown in blue.}
  \label{fig:ARM_Fit}
\end{figure}

The Compton angular resolution calibration is a measurement of how precise the instrument can determine the direction of an incoming gamma-ray photon. To calculate the angular resolution measurement (ARM), a radioactive source is placed at a known location above the instrument, and the radiated photons are measured by the detectors. Each measured photon is then reconstructed via the Compton equation (Eq.~\ref{eq:Compton}) to determine the event circle. The shortest distance from the reconstructed event circle to the known location of the source is defined as the ARM \cite{Zoglauer_2005}. The histogram of ARM values from all events, as illustrated in Fig.~\ref{fig:ARM_Fit}, usually takes on a Gaussian or Lorentzian shape, the FWHM of which is defined as the angular resolution. Theoretically, the limiting factor of angular resolution is set by Doppler broadening, since scattering electrons are bound and in motion rather than free and at rest as assumed in the Compton equation \cite{DuMond_1929}. In practice, however, it is primarily constrained by the precision of the position and energy measurements.

\begin{figure}[h]
  \centering
  \includegraphics[width=0.40\textwidth]{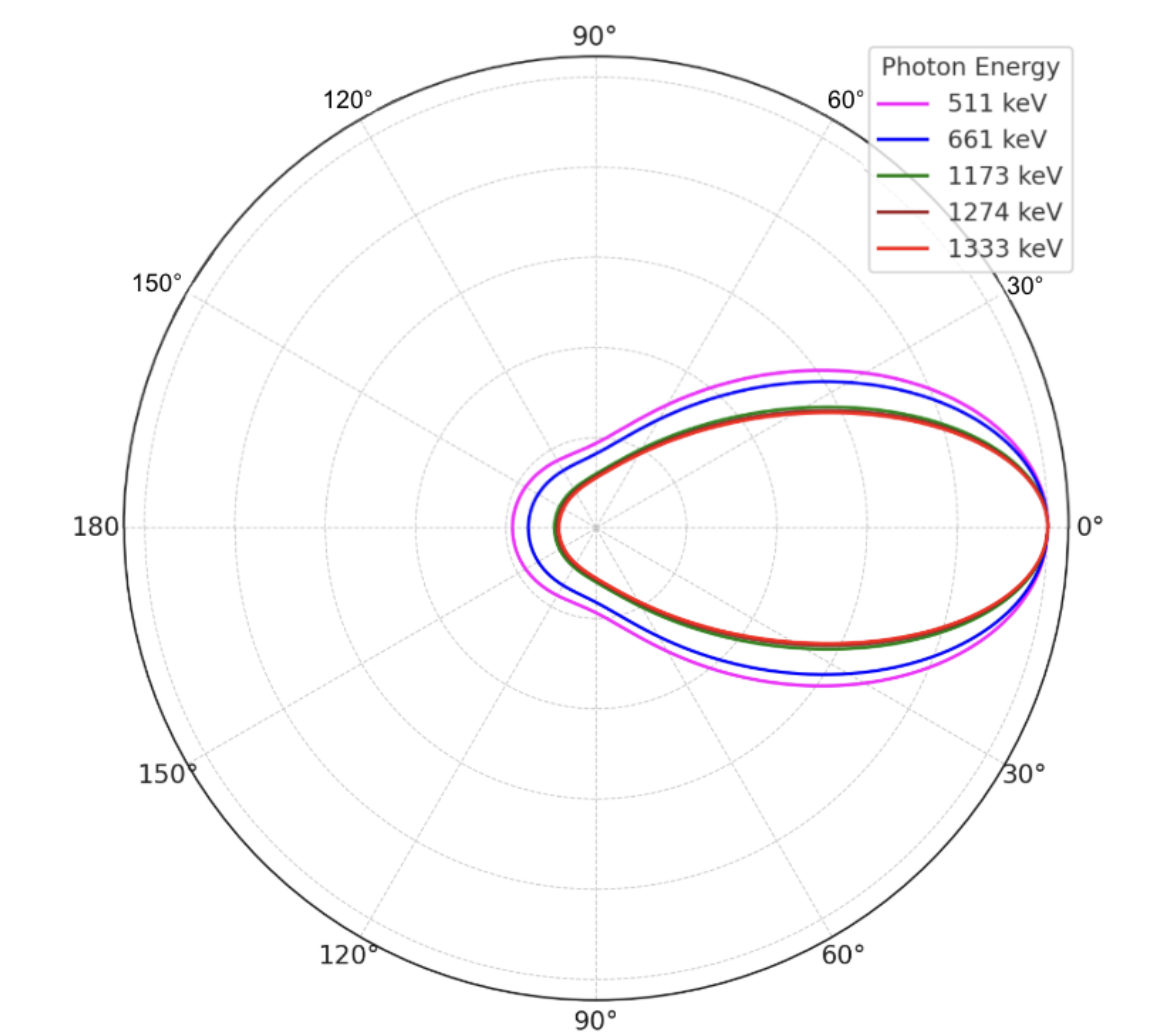}
  \caption{Klein-Nishina distribution of scattering angle cross-sections at the energies used for calibrations.}
  \label{fig:Klein-Nishina}
\end{figure}

\begin{figure}[H]
  \centering
  \begin{minipage}[t]{0.48\textwidth}
    \centering
    \includegraphics[width=\textwidth]{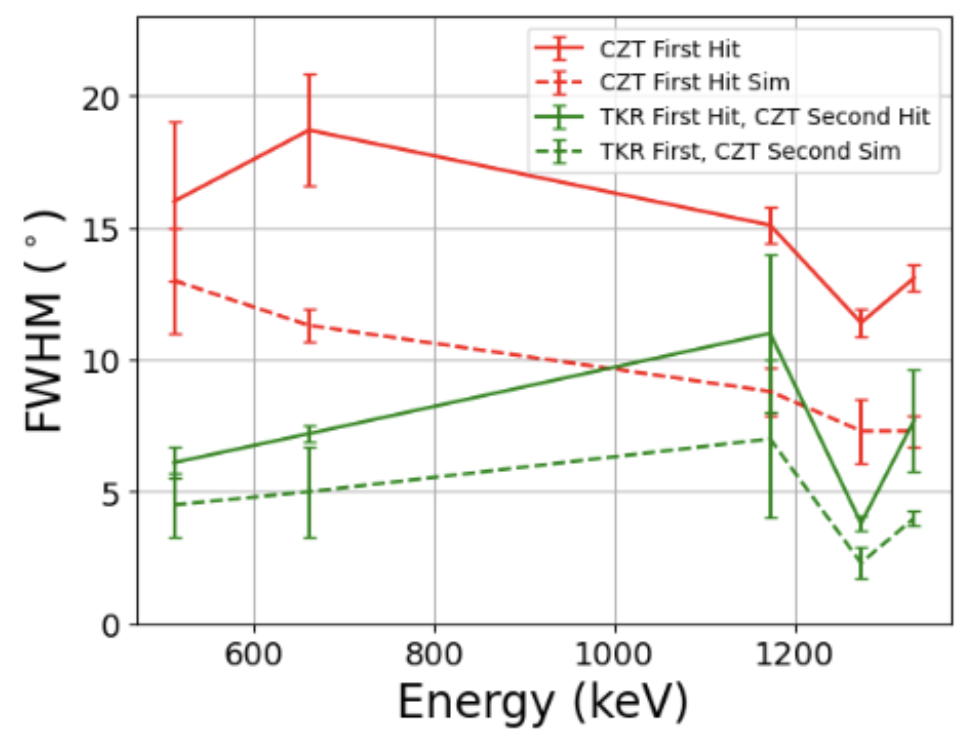}
    \caption{The FWHM of the ARM as a function of energy, separated by events that hit the Tracker first and CZT second (green), and events that hit the CZT first (red). The simulations of these two selections is shown with the dotted lines.}
    \label{fig:ARM}
  \end{minipage} \hfill
  \begin{minipage}[t]{0.48\textwidth}
    \centering
    \includegraphics[width=\textwidth]{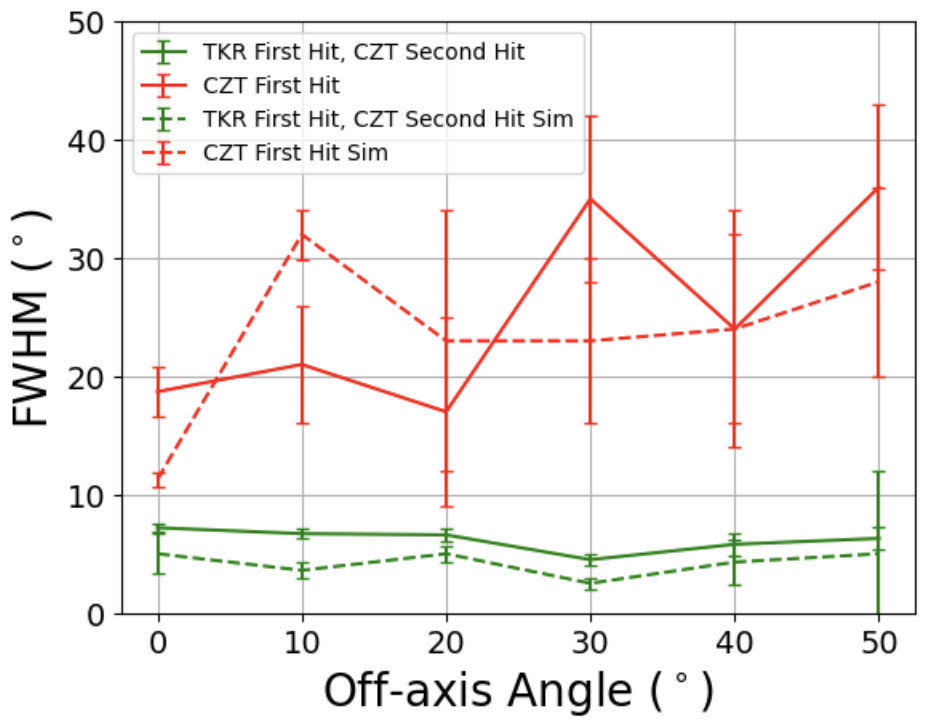}
    \caption{ARM at 661.7 keV as a function of off-axis angle, separated by events that hit the Tracker first and CZT second, and events that hit the CZT first.}
    \label{fig:off-axis_ARM}
  \end{minipage}
\end{figure}

The angular resolution was measured using an energy cut of $\pm$ 1 FWHM around the photopeak of the Compton spectrum. In the ideal case, a gamma-ray photon will Compton scatter in either the Tracker or CZT before being absorbed in one of the calorimeters. Therefore, two event selections were made. The most stringent selection, for the best angular response, required the first interaction to be in the Tracker, followed by an interaction in the CZT. Figure~\ref{fig:Klein-Nishina} shows the Klein-Nishina cross section at the energies used for calibrations, where the vast majority of Compton events scatter at angles less than 60$^\circ$. This motivated a cut requiring the Compton scattering angle to be below 60$^\circ$, in order to isolate the best events. Since the Tracker has the best position resolution of the subsystems, this selection resulted in the best angular resolution. The second selection required the first interaction to be in the CZT Calorimeter. Since the CZT cannot effectively resolve multiple hits in the same bar or bars that share a cathode readout, a cut was made that required the distance between two interactions to be greater than 2 cm. Figure~\ref{fig:ARM} shows the measured ARM FWHM as a function of energy for both of these selections, as well as simulations for the two selections. The uncertainty in the FWHM was estimated by bootstrapping the data and performing 50 repeated fits, as provided in the Mimrec ARM tool. The ARM at 662 keV for both selections was found to be 7.2$^\circ$ $\pm$ 0.3$^\circ$ and 18.7$^\circ$ $\pm$ 2.1$^\circ$, respectively.

The angular resolution is expected to improve with increasing photon energy due to several factors: a) higher-energy photons tend to scatter at smaller angles, which narrows the ARM distribution; b) the impact of Doppler broadening diminishes at higher energies; and c) the energy resolution improves. This is observed for events that interacted in the CZT first, and agree with simulations, which provide a more idealized ARM, but still maintain the overall shape. However, this trend is absent in events that begin in the Tracker. This could be partly attributed to limitations in the Tracker energy calibration, which is less reliable at high energies. 

It is noted that the ARM exhibits a weird behavior at high energy: it initially degrades from 511 keV to 1173 keV, improves at 1274 keV, and then degrades again at 1333 keV. This pattern is observed both in the measured data and in simulations. This trend, while unexpected, could be attributed to the relative prominence of the peaks against their respective backgrounds. The 1173 and 1333 keV data points, originating from $^{60}$Co, are situated on a higher background level, whereas the 1274 keV peak from $^{22}$Na stands out more distinctly against its background (see plots in \ref{app2}). However, the data points from the $^{60}$Co have large uncertainties and are still within expected variation. Additionally, events originating in the Tracker were globally more accurately reconstructed than those beginning in the CZT. This highlights the importance of the Tracker’s high position resolution for achieving precise event reconstruction in the Compton regime. 

Figure~\ref{fig:off-axis_ARM} shows the measured ARM FWHM as a function of off-axis angle for the $^{137}$Cs 662 keV line. Across the sampled angles, the ARM remained stable for events that originated in the Tracker and included a CZT interaction. This consistency is likely due to the stringent event selection, which filtered out lower-quality reconstructions and mitigated the effects of increasing off-axis angle. Additionally, this demonstrates that while the field of view starts cutting off around 20$^\circ$ (to be discussed in Section 5.3), ComPair remains capable of reliably reconstructing events at significantly larger angles. The ARM for events starting in the CZT, however, did show degradation over off-axis angle. A degradation in the ARM at higher incident angles arises from geometric effects. As the angle increases, photons are more likely to scatter into neighboring detectors as opposed to between detector planes. This results in a shorter average distance between the first two interactions. The off-axis simulations were in agreement with the real data. For events originating in the Tracker, the simulated data took on the same shape and remained constant over all sampled angles. For CZT-first events, there were observed discrepancies at lower angles; however, the simulated data still globally matched.

\begin{figure}[h]
  \centering
  \includegraphics[width=0.45\textwidth]{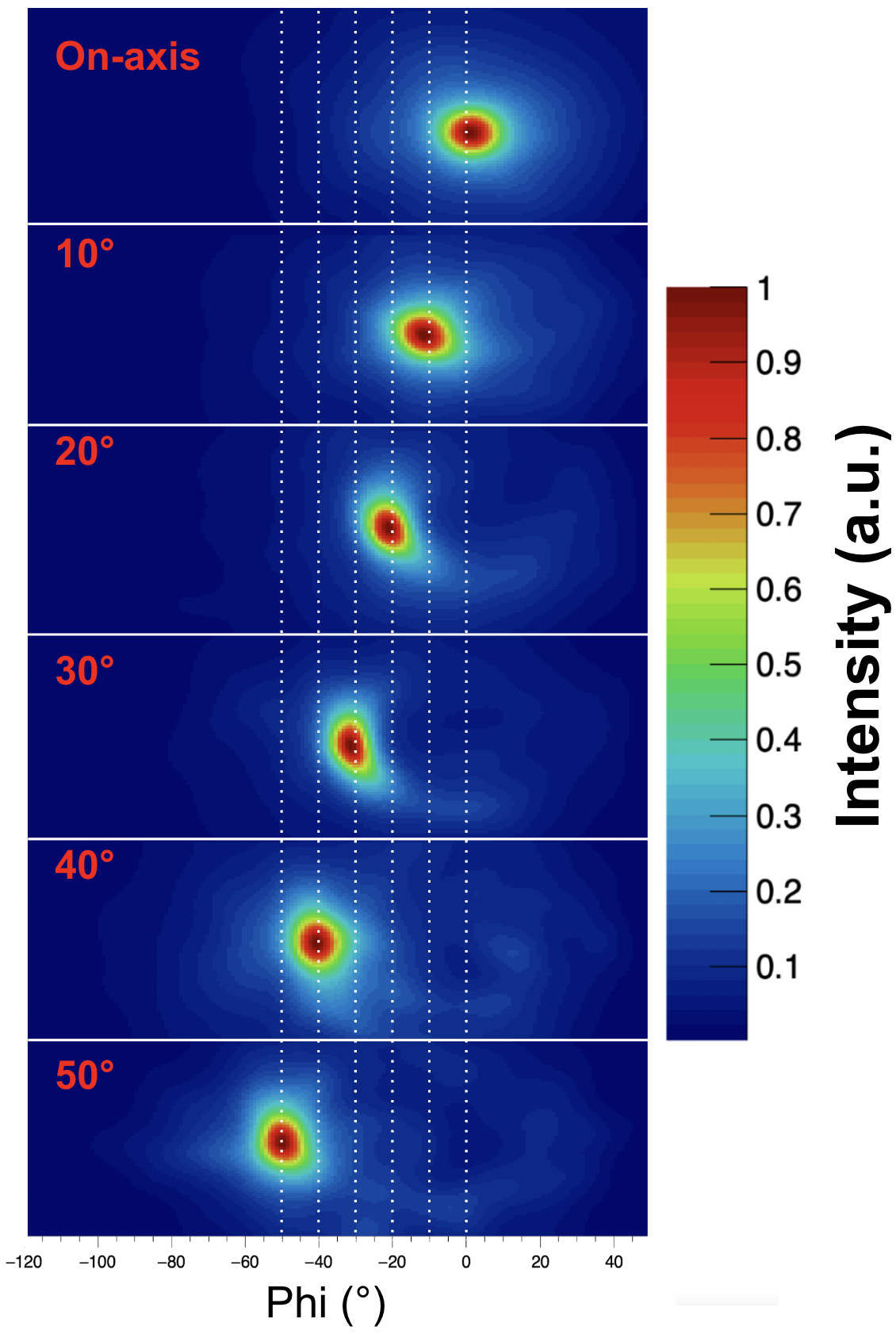}
  \caption{Reconstructed images of a $^{137}$Cs source at various angles. White dotted lines are included to guide the viewer's eye to the locations corresponding to each angle.}
  \label{fig:ReconstructedIMG}
\end{figure}

Finally, to demonstrate ComPair's Compton imaging capabilities, reconstructed images of $^{137}$Cs, generated in Mimrec, for each off-axis run is shown in Fig.~\ref{fig:ReconstructedIMG}. All six images were accurately reconstructed at their respective off-axis angles, demonstrating ComPair's reliable performance across the field of view.

\subsection{Effective Area}

The effective area determines how sensitive the instrument is to incoming photons by quantifying the rate at which photons from a source are detected. The effective area is calculated as:
\begin{equation}
    \label{eq:Effective_area}
    A_{eff} = \frac{A}{a_s \cdot P_T}\cdot4\pi d^2,
\end{equation}
where A is the rate of detected photons in the photopeak (measured in counts/s), $a_s$ is the source activity (Hz), $P_T$ is the transition probability, and $d$ is the distance between the source and the detector (measured in cm). All sources used for calibrations came with a calibration certificate from the supplier that listed the activity of the source with an uncertainty of 3\%. The current activity level was then calculated based on the amount of time elapsed to the activity calibration.

\begin{figure}[h]
  \centering
  \begin{minipage}[t]{0.48\textwidth}
    \centering
    \includegraphics[width=\textwidth]{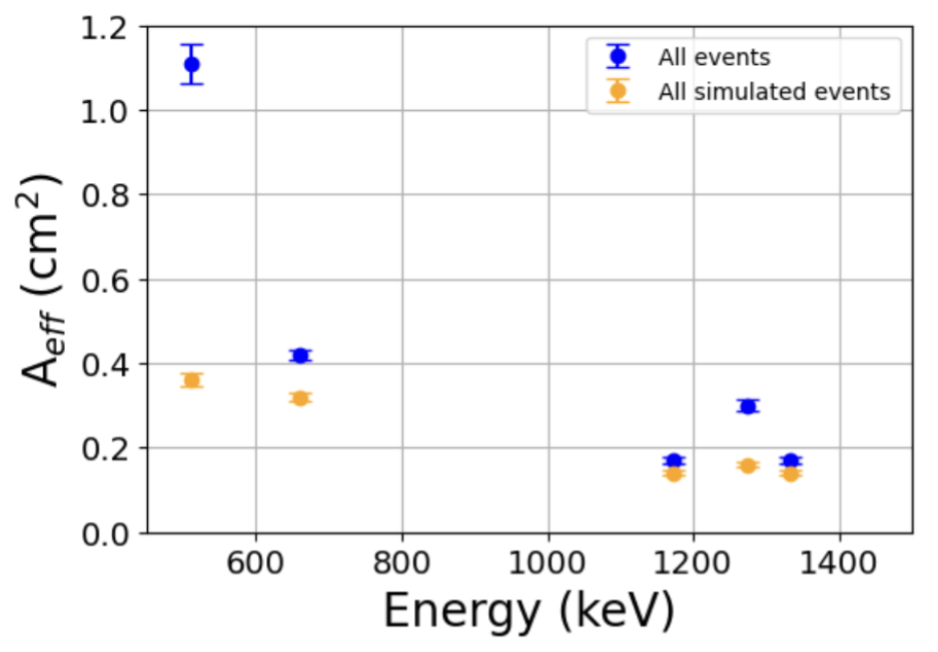}
    \caption{Effective area as a function of energy (in blue) versus the simulated effective area (in orange).}
    \label{fig:EffA}
  \end{minipage} \hfill
  \begin{minipage}[t]{0.48\textwidth}
    \centering
    \includegraphics[width=\textwidth]{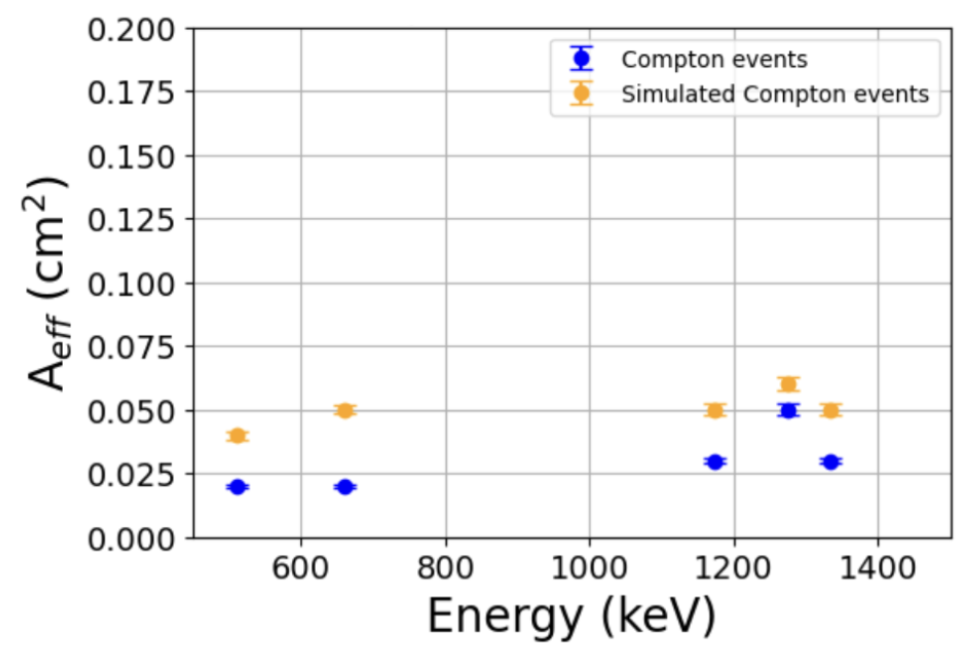}
    \caption{Compton effective area as a function of energy (in blue) versus the simulated Compton effective area (in orange).}
    \label{fig:EffA_Compton}
  \end{minipage}
\end{figure}

The rate of detected photons in the photopeak (A) was calculated by integrating the background-subtracted photopeak over the range spanning three standard deviations on either side of the photopeak’s mean, representing 99\% containment, and then dividing by the total run time (in seconds). Figures~\ref{fig:EffA} and \ref{fig:EffA_Compton} show the effective area as a function of energy, calculated for both the full instrument spectrum (including single-site events) and the Compton-only spectrum. Error bars were derived from the 3\% uncertainty in activity as well as uncertainty in the height of the source (discussed in Section 3.1). The effective area at 662 keV was found to be 0.42 cm$^2$ $\pm$ 0.01 cm$^2$ and the Compton effective area at 662 keV was found to be 0.020 cm$^2$ $\pm$ 0.001 cm$^2$.

The effective area of the full instrument decreases with energy, reflecting the declining probability of photoabsorption and the increasing probability of photons passing through the detector without depositing their full energy. This is reflected in the simulations as well. In contrast, the Compton effective area shows a slight increase with energy, consistent with the growing dominance of Compton scattering in this energy range. This trend suggests that while the overall detection efficiency drops, a larger fraction of detected events are suitable for Compton reconstruction. For the Compton effective area, the current simulations are once again in good overall agreement with the measured data, accurately capturing the shape of the distribution while slightly overestimating its magnitude. However, for the full instrument effective area, the simulations underestimated the data, and a large discrepancy is noted at 511 keV which is still being investigated. Various adjustments, including modifying the simulated subsystem thresholds, have been tested to better match the simulation with the data, however, this data point still remains an outlier.

\begin{figure}[h]
  \centering
  \includegraphics[width=0.45\textwidth]{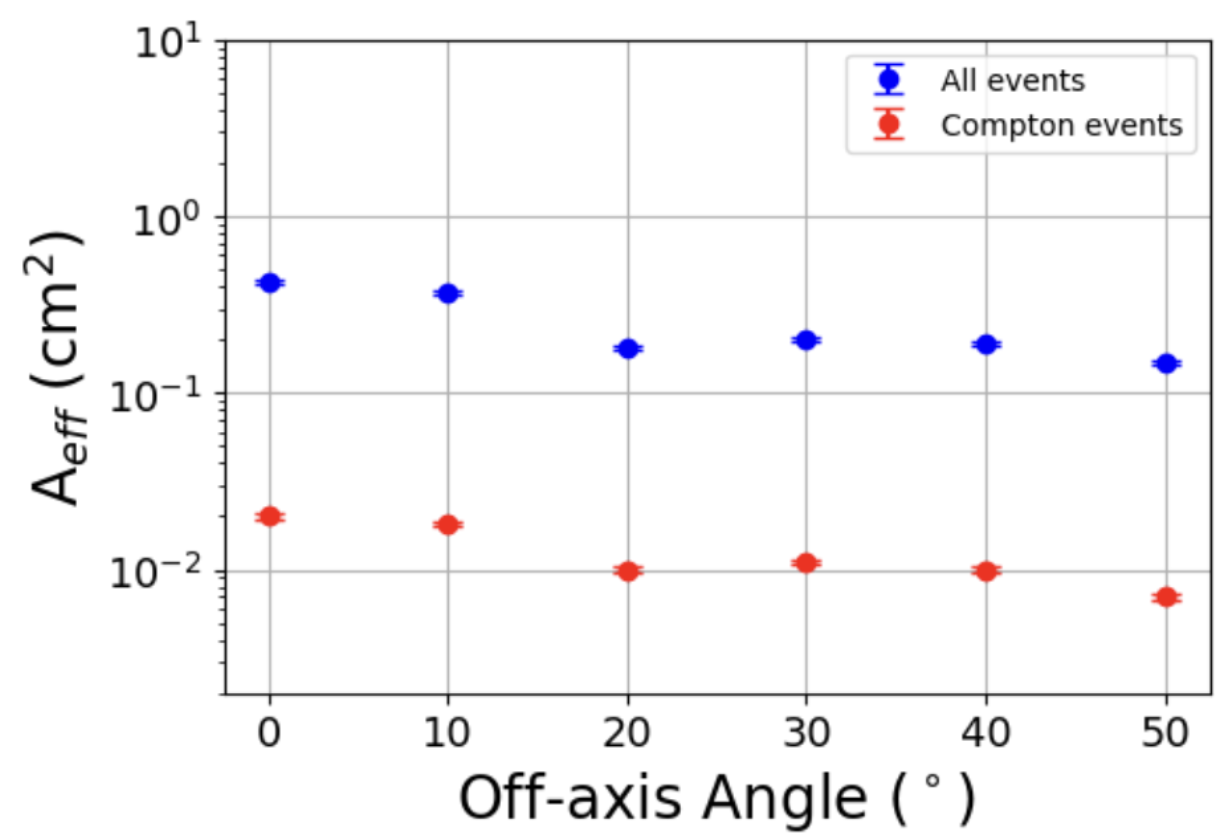}
  \caption{Effective area as a function of off-axis angle for all events (blue) and Compton events (red).}
  \label{fig:off-axis_EffA}
\end{figure}

Additionally, the effective area was measured as a function of off-axis angle using a $^{137}$Cs source. This was done for all events and Compton-only events, the results of which are shown in Fig.~\ref{fig:off-axis_EffA}. As the off-axis angle increases, the geometric overlap between the Tracker and the calorimeters diminishes, reducing the likelihood of coincident interactions. Since successful event detection requires such coincidences, the effective area correspondingly decreases. Additionally, at higher angles, the incoming particles will have to traverse through more material, which increases the chances of energy loss. Therefore one would expect the effective area to drop as a function of off-axis angle.

To better understand the low efficiency, we track where events are lost in the pipeline, summarized in Table~\ref{tab:Events_Loss}. This not only will offer an explanation for why the effective area is admittedly poor but can also offer insight on ways to improve for future generations.

For the on-axis $^{137}$Cs run, we started with 4.34 million events collected over the course of 12 hours. 3.74 million of these events measured above threshold for all 4 pads in the CZT pipeline (see section 4.2) and therefore passed selection, with 3.02 million events then passing the CZT top-of-bar cut. 2.84 million of these events passed Revan's event selections, while $\sim$181,000 were deemed unreconstructable. The majority of these rejected events failed due to either the event consisting of nothing but one track (85\%), or no good Compton sequence reconstruction combination was found (11\%). Of the events that did pass however, 84.4\% were identified as single-site events (mostly photoabsorption events but can also be from incomplete absorption) and 8.7\% were identified as Compton events. Finally, $\sim$577,000 of the 2.84 million events were within the photopeak (99\% containment), with $\sim$341,000 remaining after background subtraction.

\begin{table}[ht]
\centering
\resizebox{0.45\textwidth}{!}{%
\begin{tabular}{lrrr}
\hline
\textbf{Stage} & \textbf{Remaining} & \textbf{\% of Previous} & \textbf{\% of Initial} \\
\hline
Initial collected events & 4,340,000 & 100.0\% & 100.0\% \\
CZT pad cut passed & 3,740,000 & 86.2\% & 86.2\% \\
CZT top-of-bar cut passed & 3,020,000 & 80.7\% & 69.6\% \\
Reconstructable events$^\dagger$ & 2,840,000 & 94.0\% & 65.4\% \\
\quad Single-site events & 2,396,960 & 84.4\% & 55.2\% \\
\quad Compton events & 247,080 & 8.7\% & 5.7\% \\
Within photopeak & 577,000 & 20.3\% & 13.3\% \\
Background subtracted & 341,000 & 59.1\% & 7.9\% \\
\hline
\end{tabular}
}
\vspace{0.5em} \\
\footnotesize{
$^\dagger$ Percentage of Revan-passed events.
}
\caption{Event retention and loss at each stage of the on-axis $^{137}$Cs run. Percentages are relative to the previous step and to the total initial events (4.34 million).}
\label{tab:Events_Loss}
\end{table}

Additionally, the event loss rate can be examined in simulations. For a $^{22}$Na source simulations, 3,213,000 events in the raw simulation passed Revan reconstruction. After running the simulation through the DEE however, only 397,000 passed Revan reconstruction. This shows how big an impact subsystem detection thresholds and trigger efficiencies have on ComPair's overall performance.

It is acknowledged that ComPair had a relatively low effective area which is not representative of AMEGO's performance. We performed simulations with the AMEGO mass model to understand the impact of this measured efficiency on the mission requirements. Accounting for the active area differences (AMEGO's design has $\sim64$ times the active detector area and $6$ times the number of Tracker layers as ComPair) gives a factor of $\sim384$ increase in the effective area. Additionally, the increase in collecting area for Compton scatters based on the optimized geometry gives an additional factor of $\sim10$. With these simplified geometric approximations, the expected AMEGO Compton effective area at 662 keV would be $\sim96$ cm$^2$. If the energy thresholds for each detector system meet the AMEGO requirement level, this gives a factor of $\sim2$ increase in the effective area, resulting in a predicted 190 cm$^2$ effective area at 662 keV. This is still about an order of magnitude below the expected efficiency of AMEGO; however, the additional losses from trigger inefficiencies (discussed in \cite{metzler2024}), the 2 cm distance cut, and significant passive material in the ComPair detector region can likely explain this deviation. Therefore, the low effective area of ComPair matches with our expectations based on the small active area and non-optimal detector geometry, high energy thresholds, and measured trigger inefficiencies.

\section{Conclusion}

Calibrations are essential in understanding the precision of an instrument's measurements. Through calibrations, one can characterize key performance metrics such as energy resolution, angular resolution, and effective area, all of which are critical for reconstructing gamma-ray events. Additionally, calibrations play a key role in benchmarking simulations, ensuring that modeled detector performance aligns with real-world measurements. Table. \ref{tab:Summary_results} provides a summary of the calibration results at 662 keV.

\begin{table}[h]
    \centering
    \begin{tabular}{| c | c |}
        \hline
        Energy Resolution (FWHM) &  28.57 keV $\pm$ 0.3 keV  \\ \hline 
        Compton Energy Resolution (FWHM) & 34.24 keV $\pm$ 0.1 keV  \\ \hline
        \makecell{Angular Resolution \\ (TKR First, CZT Second)} & 7.2$^\circ$ $\pm$ 0.3$^\circ$  \\ \hline
        \makecell{Angular Resolution \\ (CZT First)} & 18.7$^\circ$ $\pm$ 2.1$^\circ$  \\ \hline
        Effective Area & 0.42 cm$^2$ $\pm$ 0.01 cm$^2$  \\ \hline
        Compton Effective Area & 0.020 cm$^2$ $\pm$ 0.001 cm$^2$ \\ \hline
    \end{tabular}
    \caption{Summary of calibration results at 662 keV.}
    \label{tab:Summary_results}
\end{table}

Comparing the calibration results with simulations shows good preliminary agreement between the two, with further refinements to the DEE currently in progress to improve this agreement. Specifically, at the time of this work, the CZT reverse pipeline is still in the early stages of development. Efforts are currently focused on improving the agreement between simulated and real CZT data. The discrepancies are primarily due to challenges in handling events in shared cathodes and accurately resolving event positions when two interactions occur in adjacent bars in the real detector. Additionally, efforts are ongoing to refine the simulated subsystem energy thresholds to more accurately reflect the behavior of the real instrument.

These calibration results will also serve as a basis for comparison with ComPair-2, which is currently in development \cite{Caputo_2024}. ComPair-2 is a prototype of the AMEGO-X mission concept, the Medium Explorer version of AMEGO \cite{Caputo_2022}. ComPair-2 will be composed of a 10 layer silicon Tracker and a 4 layer CsI Calorimeter. The Tracker will feature pixelated silicon detectors with the readout electronics integrated into the detecting material. This will not only reduce the noise in the detector, but will also drastically reduce the amount of passive material. Overall, ComPair-2 will offer significantly enhanced sensitivity over its predecessor and will be designed for a long-duration balloon flight.

Within the context of AMEGO, ComPair served to validate the underlying technology while also identifying areas for improvement. For example, the combination of ComPair's small detectors with a significant amount of passive material limited its effective area. However, this limitation does not reflect the anticipated performance of AMEGO, which will have a much wider field of view, a proposed effective area \textgreater 1000 cm$^2$ at 1 MeV, and much more stopping power \cite{AMEGO}. 

This work presented the comprehensive calibration measurements of the ComPair prototype. The energy response, angular resolution, and effective area were all validated through simulations, and will provide a critical foundation for future analysis. While these results highlighted areas of strong performance, they also offered key insights for improving future gamma-ray missions such as AMEGO.

\section{CRediT authorship contribution statement}

\textbf{Nicholas Kirschner:} Methodology, Software, Validation, Formal analysis, Writing- Original Draft, Investigation, Data Curation, Writing- Review and Editing, Visualization. \textbf{Zachary Metzler:} Methodology, Software, Formal Analysis, Investigation, Data Curation, Writing- Review and Editing. \textbf{Lucas Smith}: Methodology, Software, Formal Analysis, Investigation, Data Curation, Writing- Review and Editing. \textbf{Carolyn Kierans:} Methodology, Investigation, Writing- Review and Editing, Supervision, Project Administration, Funding Acquisition. \textbf{Regina Caputo:} Methodology, Investigation, Supervision, Project Administration. \textbf{Nicholas Cannady:} Methodology, Software, Investigation. \textbf{Makoto Sasaki:} Methodology, Software, Investigation. \textbf{Daniel Shy:} Methodology, Software, Formal Analysis, Investigation, Data Curation, Writing- Review and Editing. \textbf{Priyarshini Ghosh:} Software. \textbf{Sean Griffin:} Software, Investigation. \textbf{J. Eric Grove:} Conceptualization, Funding Acquisition. \textbf{Elizabeth Hays:} Conceptualization. \textbf{Iker Liceaga-Indart:} Investigation. \textbf{Emily Kong:} Investigation. \textbf{Julie McEnery:} Conceptualization. \textbf{John Mitchell:} Investigation. \textbf{A. A. Moiseev:} Conceptualization, Methodology, Investigation, Writing- Review and Editing, Supervision. \textbf{Lucas Parker:} Software, Investigation. \textbf{Jeremy S. Perkins:} Conceptualization. \textbf{Bernard Phlips:} Conceptualization. \textbf{Adam J. Schoenwald:} Investigation. \textbf{Clio Sleator:} Investigation. \textbf{Jacob Smith:} Investigation. \textbf{Janeth Valverde:} Investigation. \textbf{Sambid Wasti:} Software. \textbf{Richard Woolf:} Methodology, Investigation, Supervision, Funding Acquisition. \textbf{Eric Wulf:} Conceptualization. \textbf{Anna Zajczyk:} Software. All authors have read and agreed to the published version of the manuscript.

\section{Data Availability Statement:} The raw data supporting the conclusions of this article will be made available by the authors on request.

\section{Declaration of competing interest}

The authors declare that they have no known competing financial
interests or personal relationships that could have appeared to influence
the work reported in this paper.

\section{Acknowledgments}

Software: matplotlib \cite{matplotlib}, MEGAlib \cite{ZOGLAUER_2006}, numpy \cite{numpy}, scipy \cite{scipy}.

\section{Funding}
This work is supported under NASA Astrophysics Research and Analysis (APRA) grants NNH14ZDA001N-APRA, NNH15ZDA001N-APRA, NNH1870A001N-APRA, and NNH21ZDA001N-APRA. The material is based upon work supported by NASA under award number 80GSFC21M0002. Research presented in this proceeding was supported by the Laboratory Directed Research and Development program of Los Alamos National Laboratory under project number 20210675ECR.

\onecolumn
\appendix
\section{Calibration Spectra}
\label{app1}

\begin{figure}[H]
    \centering

    \hspace*{-0.6cm}
    \vspace{-2cm}
    \begin{subfigure}[t]{0.47\textwidth}
        \includegraphics[width=\linewidth]{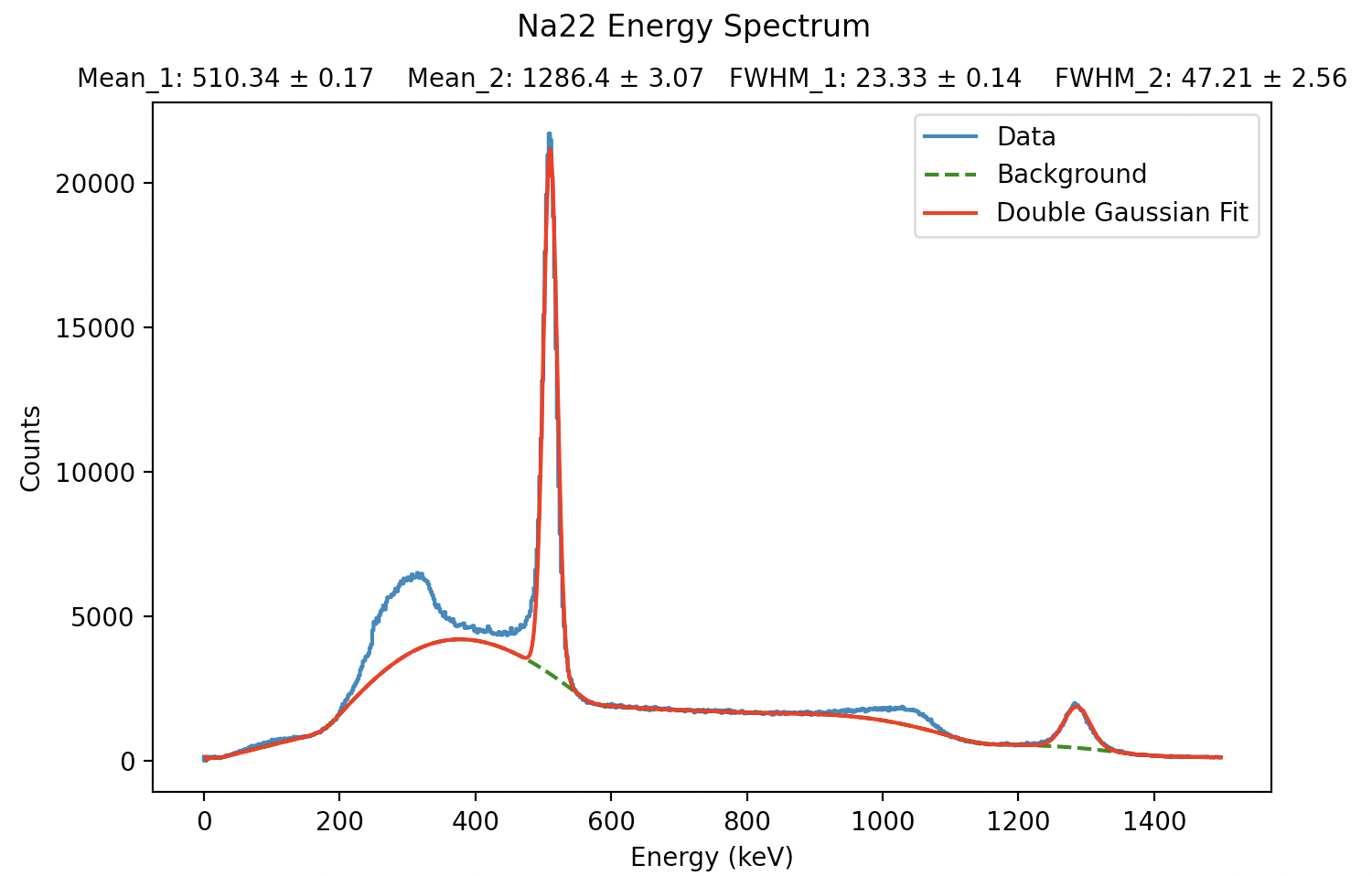}
        \caption{$^{22}$Na Spectrum}
    \end{subfigure}
    \vspace{2cm}
    \hspace*{0.1cm}
    \begin{subfigure}[t]{0.46\textwidth}
        \includegraphics[width=\linewidth]{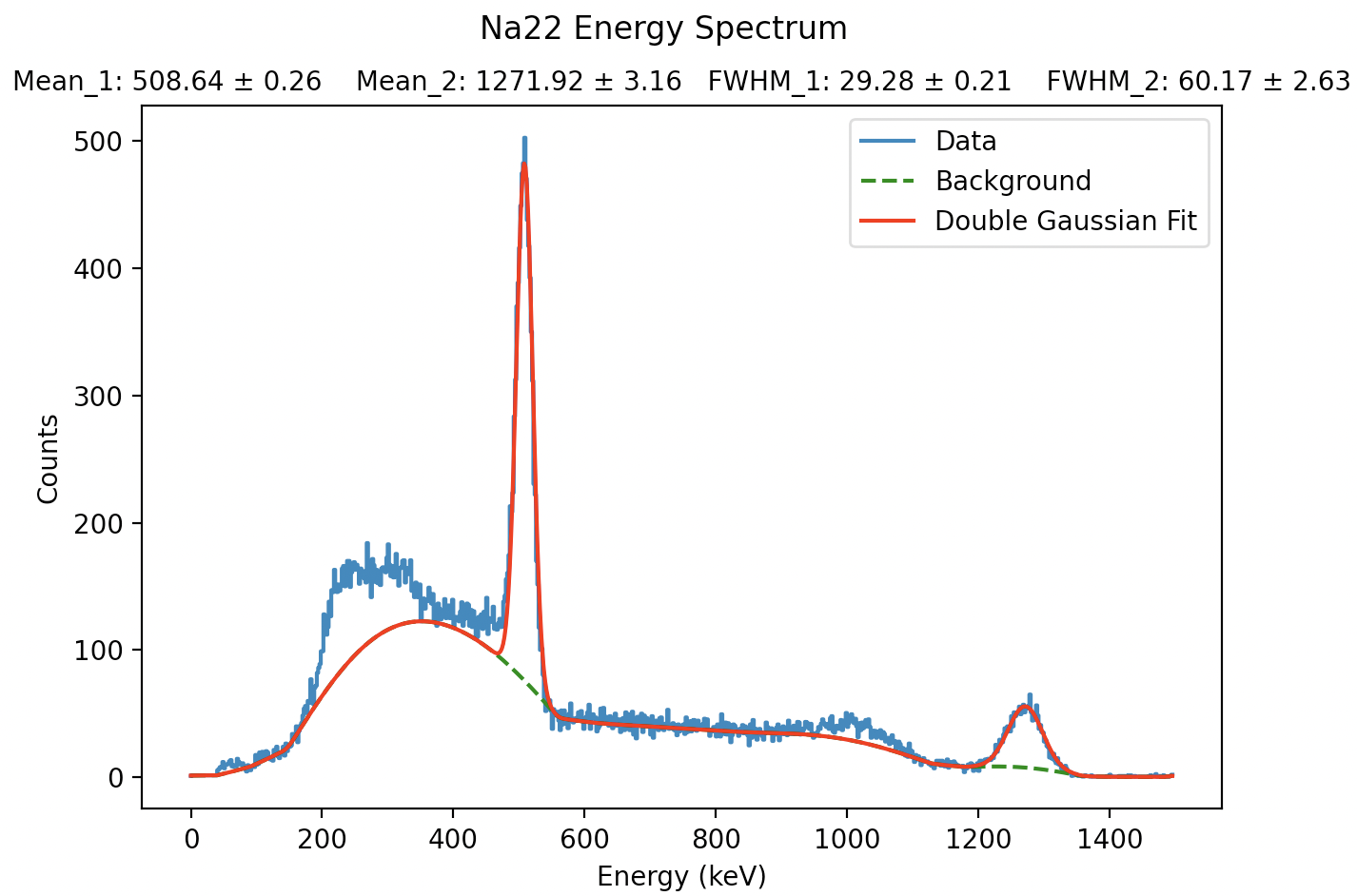}
        \caption{Simulated $^{22}$Na Spectrum}
    \end{subfigure}

    \vspace{0.5cm} 

    \hspace*{-1cm}
    \begin{subfigure}[t]{0.44\textwidth}
        \includegraphics[width=\linewidth]{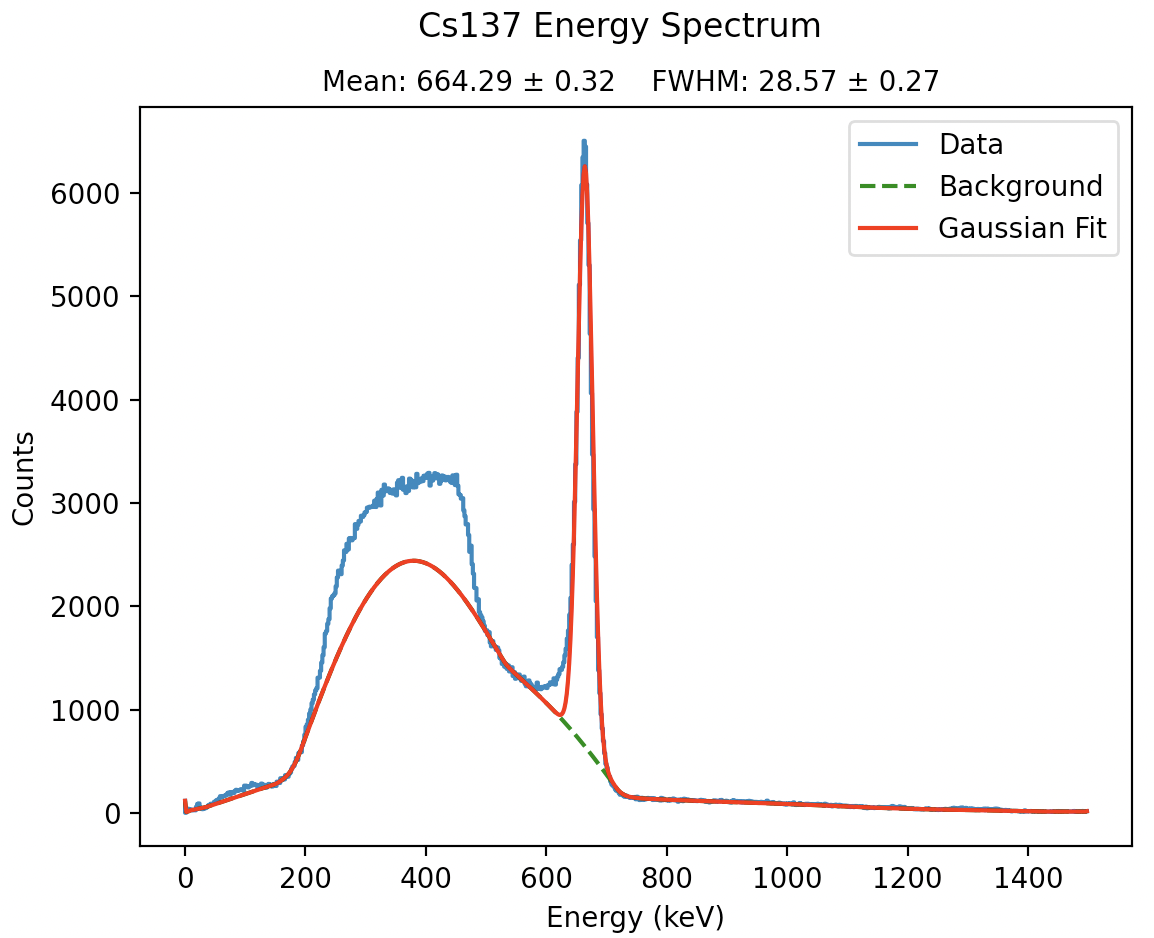}
        \caption{$^{137}$Cs Spectrum}
    \end{subfigure}
    \hspace*{0.8cm}
    \begin{subfigure}[t]{0.44\textwidth}
        \includegraphics[width=\linewidth]{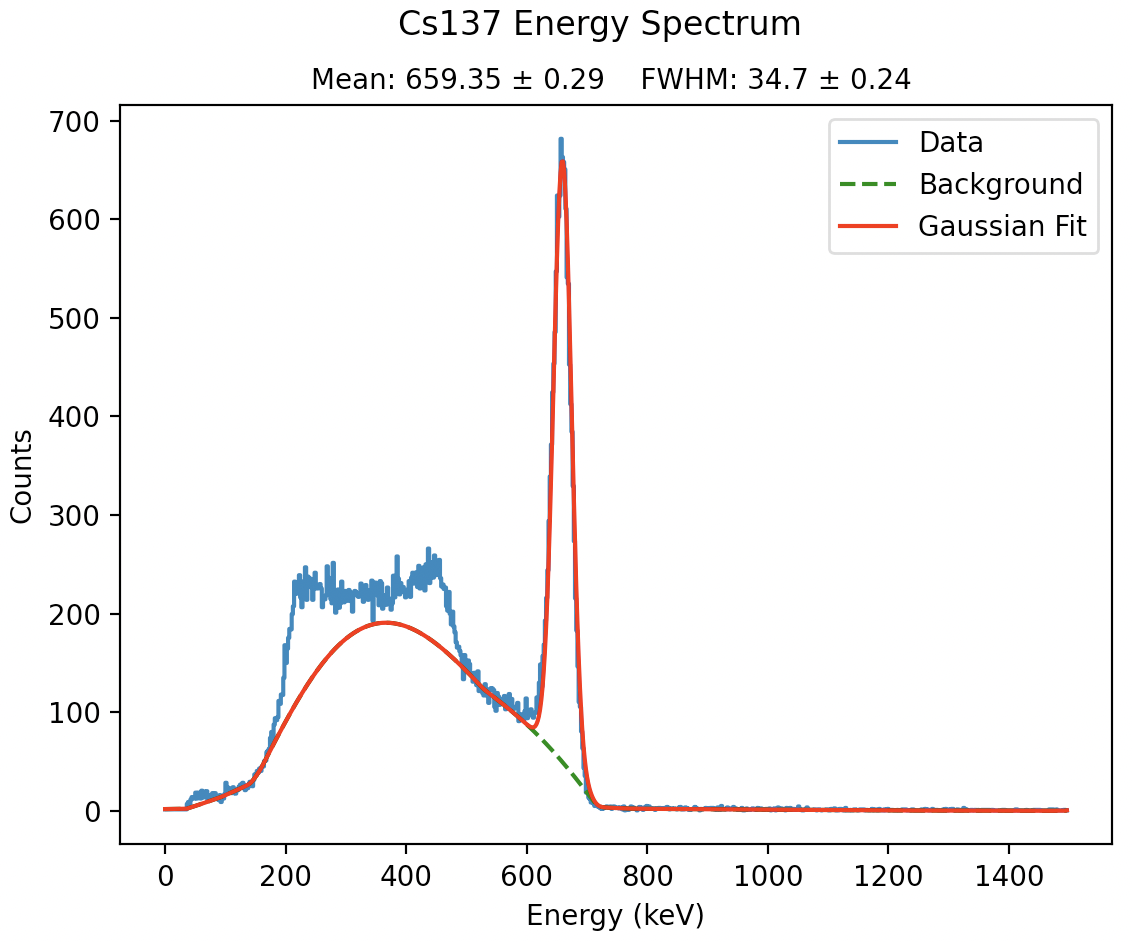}
        \caption{Simulated $^{137}$Cs Spectrum}
    \end{subfigure}

    \vspace{0.5cm}

    \hspace*{-0.5cm}
    \begin{subfigure}[t]{0.47\textwidth}
        \includegraphics[width=\linewidth]{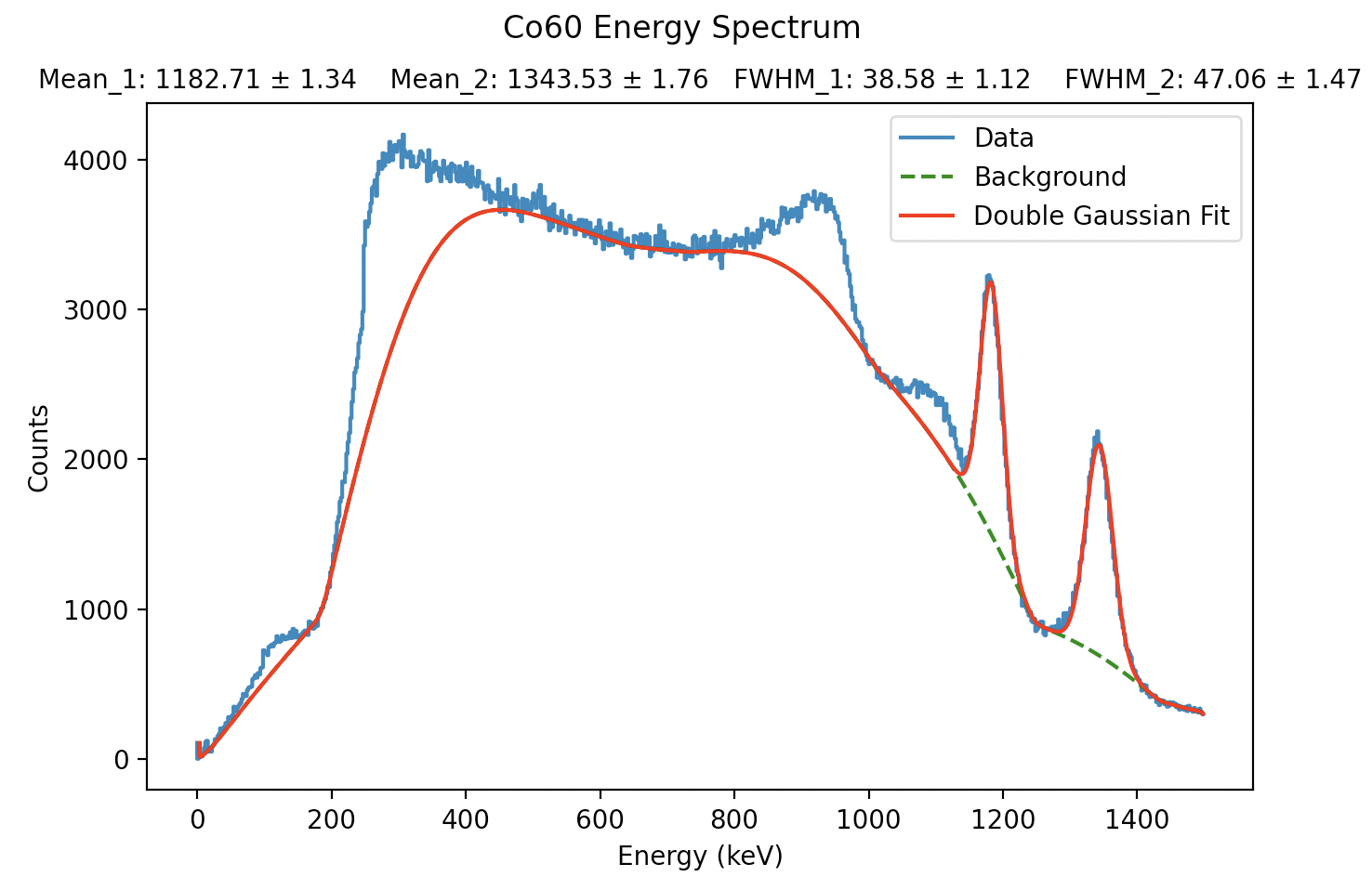}
        \caption{$^{60}$Co Spectrum}
    \end{subfigure}
    \hspace*{0.1cm}
    \begin{subfigure}[t]{0.47\textwidth}
        \includegraphics[width=\linewidth]{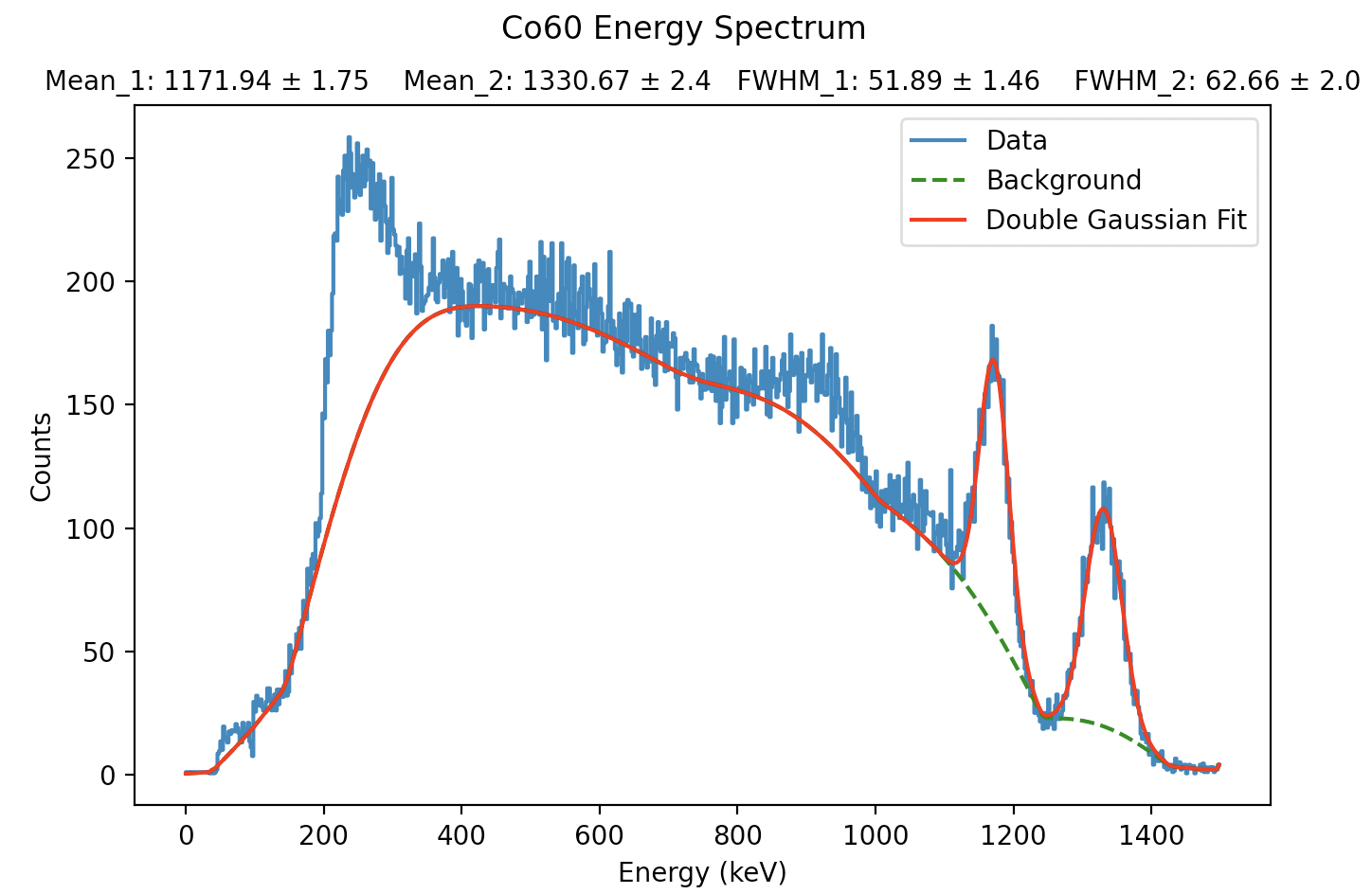}
        \caption{Simulated $^{60}$Co Spectrum}
    \end{subfigure}

    \caption{Comparison of real (left) and simulated (right) spectra from 3 radioactive sources. The spectra are fit with Gaussian or Double-Gaussian functions to obtain the FWHM of the photopeaks.}
    \label{fig:spectra-comparison}
\end{figure}

\newpage
\section{Compton Calibration Spectra}
\label{app2}

\begin{figure}[H]
    \centering

    \hspace*{-0.6cm}
    \vspace{-2cm}
    \begin{subfigure}[t]{0.47\textwidth}
        \includegraphics[width=\linewidth]{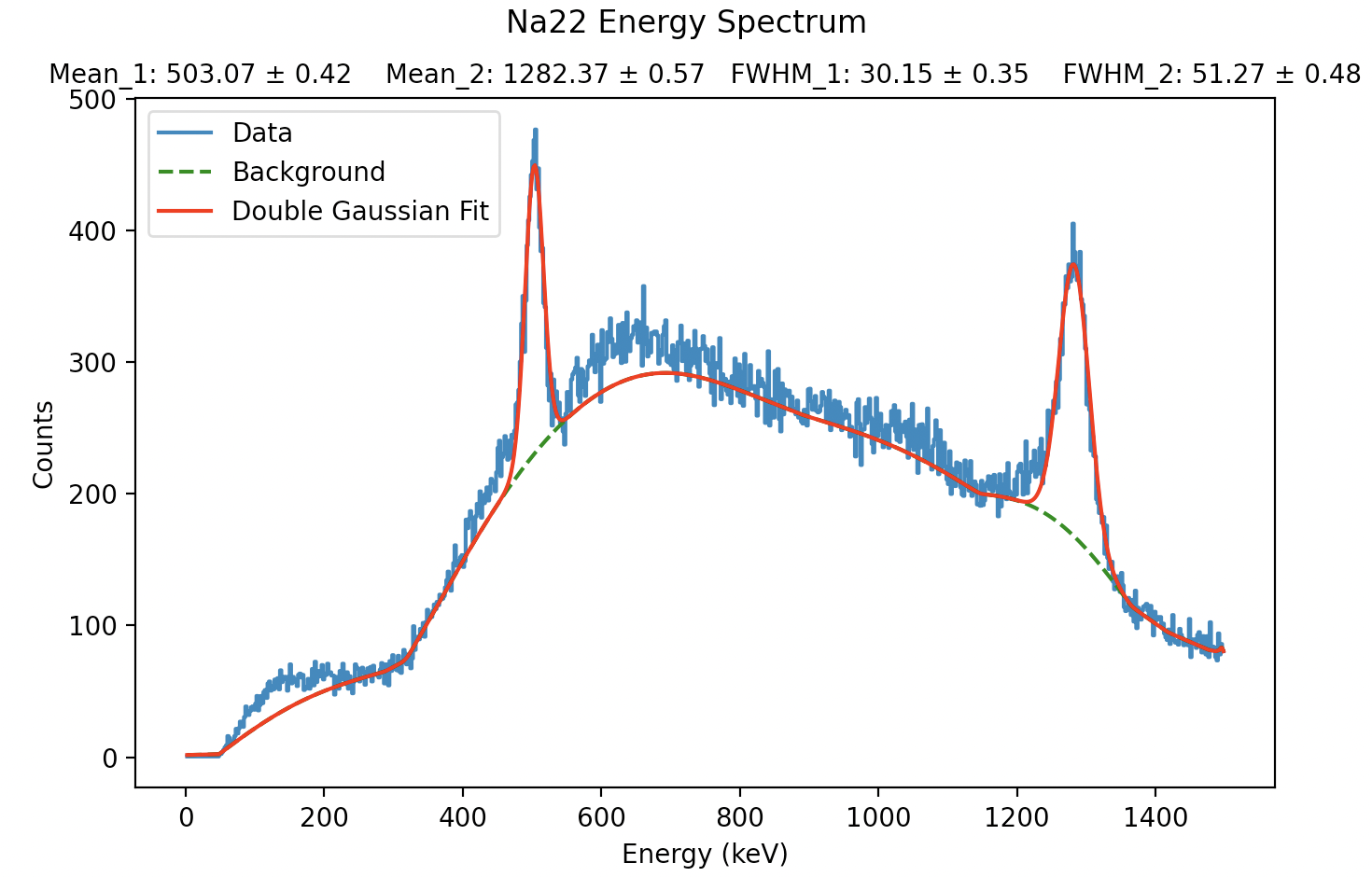}
        \caption{$^{22}$Na Compton Spectrum}
    \end{subfigure}
    \vspace{2cm}
    \hspace*{0.1cm}
    \begin{subfigure}[t]{0.47\textwidth}
        \includegraphics[width=\linewidth]{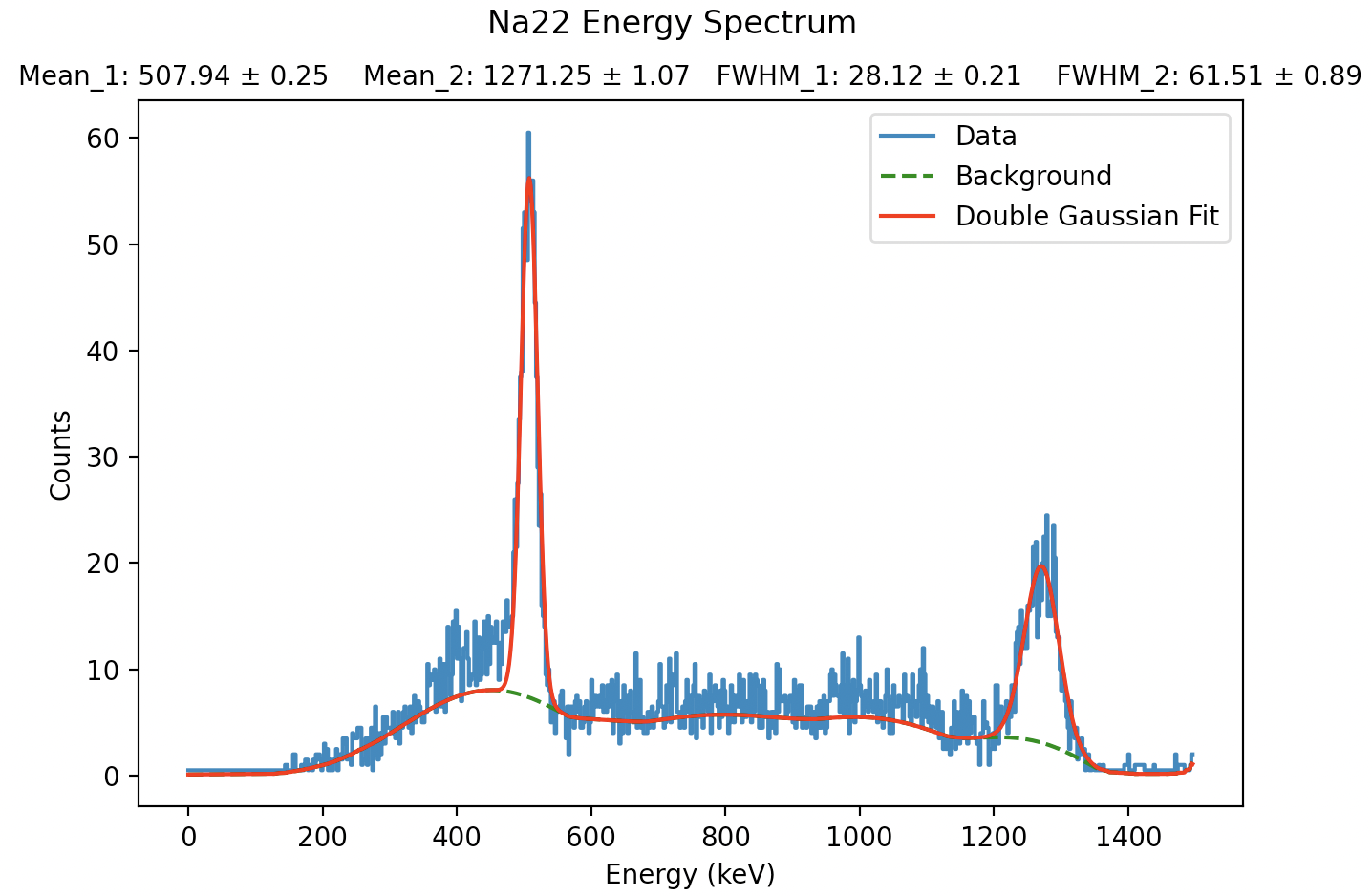}
        \caption{Simulated $^{22}$Na Compton Spectrum}
    \end{subfigure}

    \vspace{0.5cm} 

    \hspace*{-1cm}
    \begin{subfigure}[t]{0.45\textwidth}
        \includegraphics[width=\linewidth]{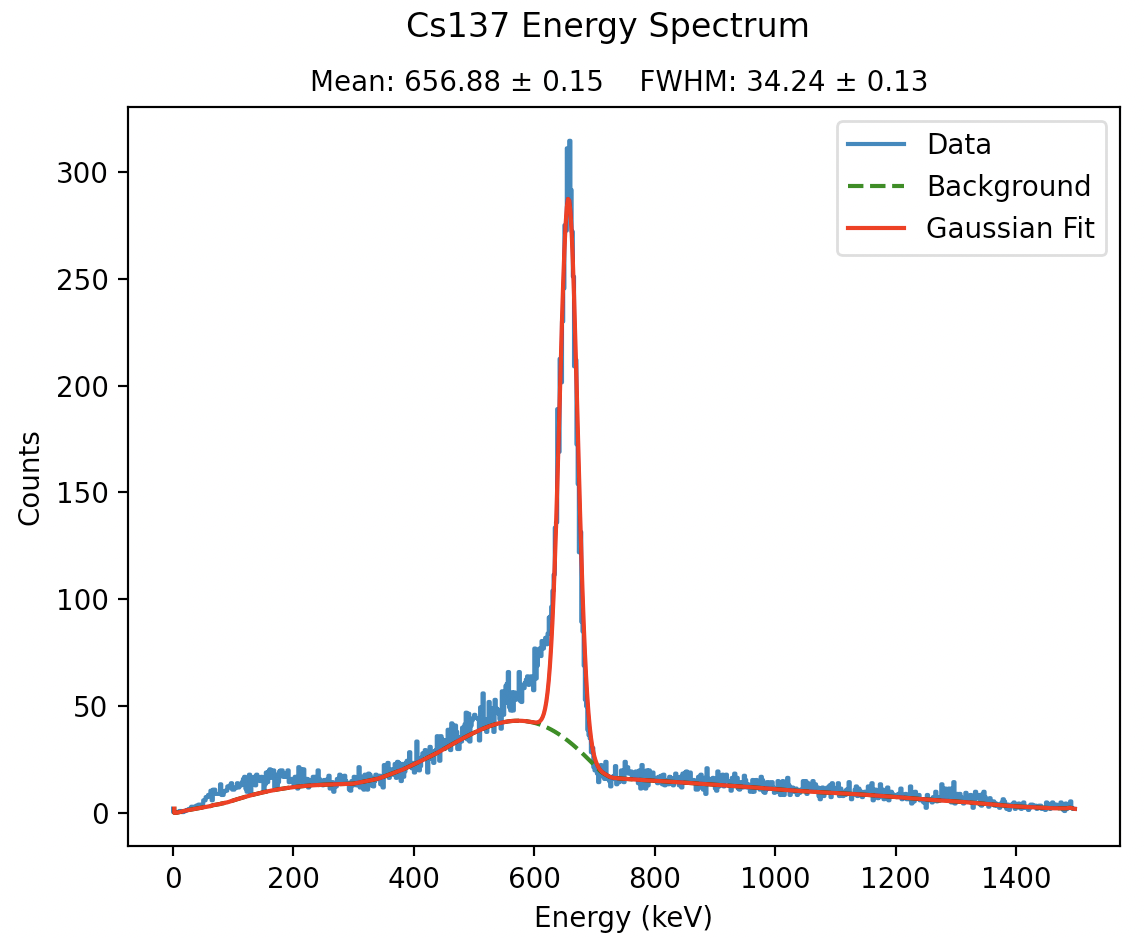}
        \caption{$^{137}$Cs Compton Spectrum}
    \end{subfigure}
    \hspace*{0.8cm}
    \begin{subfigure}[t]{0.45\textwidth}
        \includegraphics[width=\linewidth]{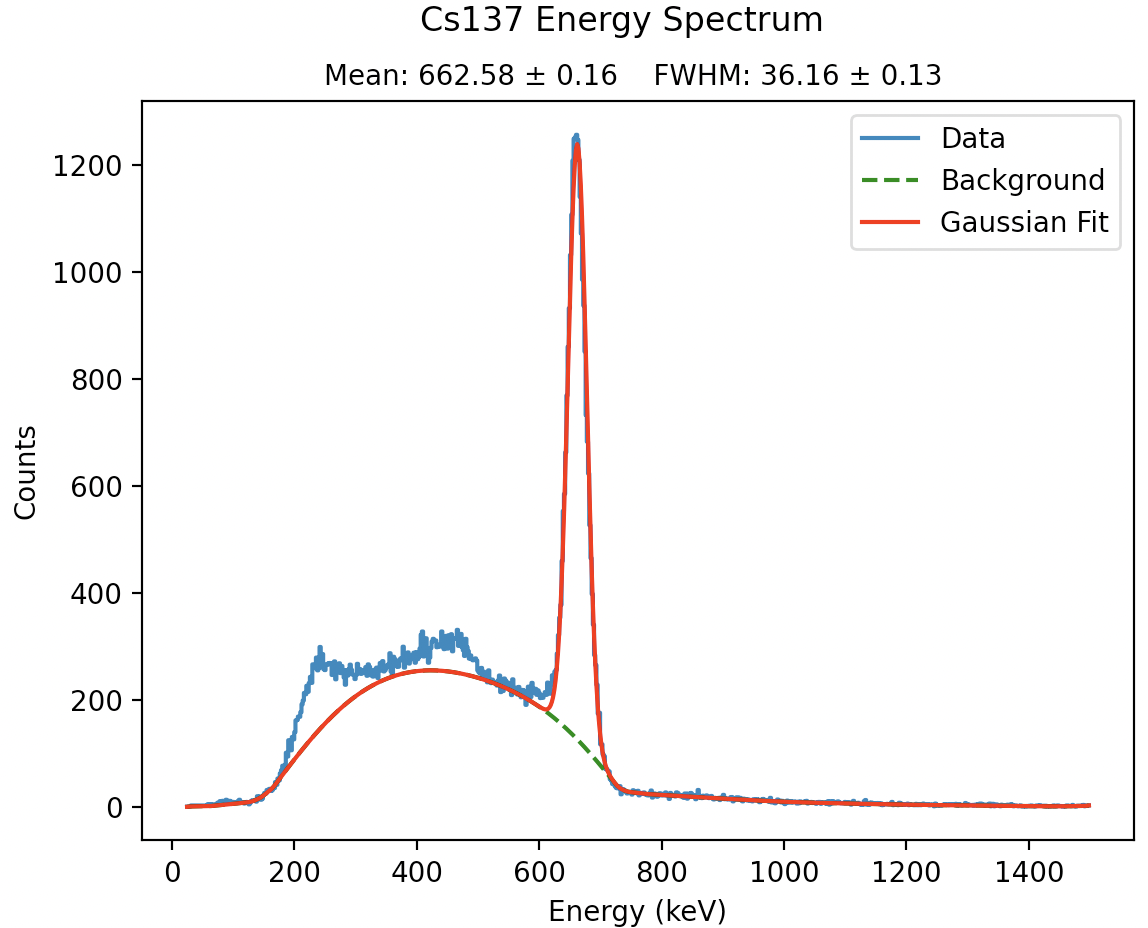}
        \caption{Simulated $^{137}$Cs Compton Spectrum}
    \end{subfigure}

    \vspace{0.5cm}

    \hspace*{-0.5cm}
    \begin{subfigure}[t]{0.47\textwidth}
        \includegraphics[width=\linewidth]{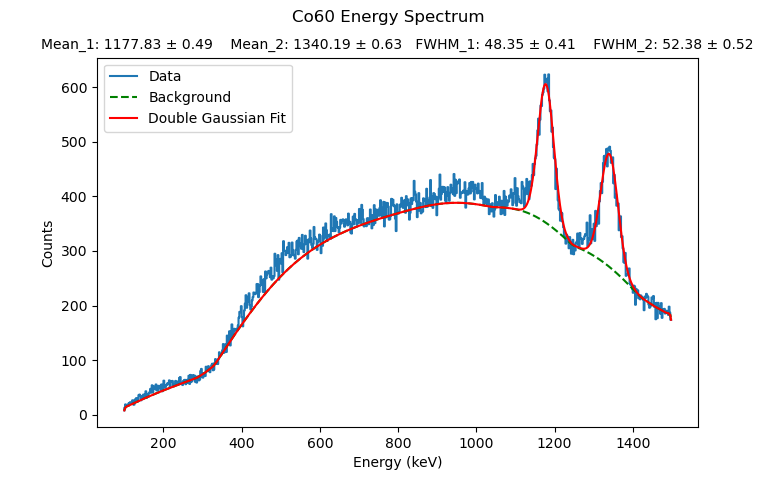}
        \caption{$^{60}$Co Compton Spectrum}
    \end{subfigure}
    \hspace*{0.1cm}
    \begin{subfigure}[t]{0.47\textwidth}
        \includegraphics[width=\linewidth]{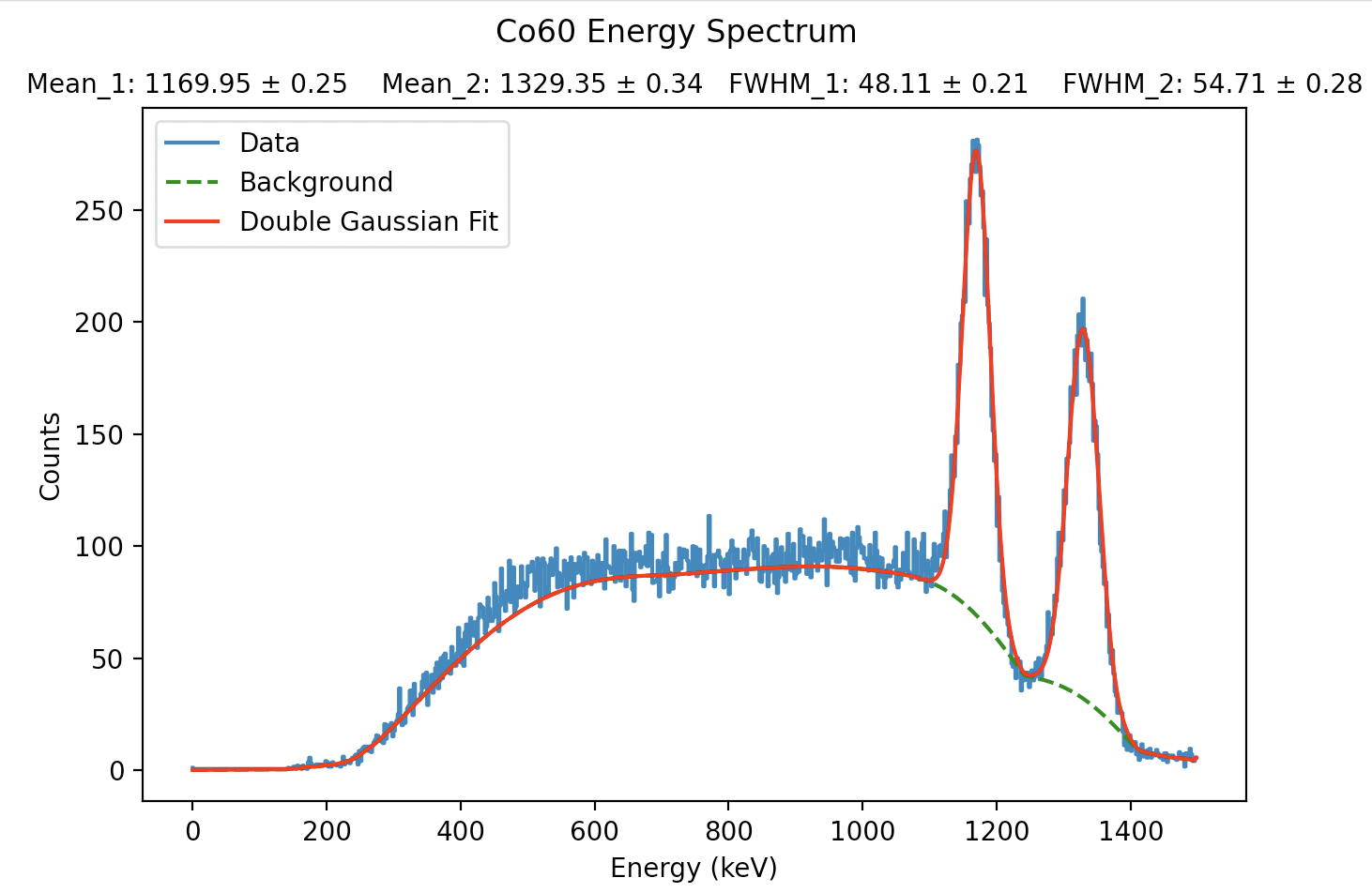}
        \caption{Simulated $^{60}$Co Compton Spectrum}
    \end{subfigure}

    \caption{Comparison of real (left) and simulated (right) Compton spectra from 3 radioactive sources. The spectra are fit with Gaussian or Double-Gaussian functions to obtain the FWHM of the photopeaks.}
    \label{fig:Compton_spectra-comparison}
\end{figure}



\newpage
\bibliographystyle{unsrt}
\bibliography{report}



\end{document}